\RequirePackage{fix-cm}
\documentclass[smallextended]{svjour3}       
\smartqed  
\usepackage{amsmath}
\usepackage{amssymb}
\usepackage{graphicx}
\usepackage{natbib}
\usepackage{url} 
\usepackage[shortlabels]{enumitem}
\usepackage[ruled]{algorithm2e}
\usepackage{dsfont,bm}
\newcommand\independent{\protect\mathpalette{\protect\independenT}{\perp}}
\def\independenT#1#2{\mathrel{\rlap{$#1#2$}\mkern2mu{#1#2}}}

\usepackage[font={small}]{caption}
\usepackage{xcolor}
\definecolor{linkcolor}{rgb}{0,0.2,0.6}
\usepackage[colorlinks,breaklinks,citecolor=black,urlcolor=linkcolor,linkcolor=black]{hyperref}

\begin{document}

\title{Testing Conditional Independence in \\ Supervised Learning Algorithms 
}


\author{David S. Watson\textsuperscript{1*} \and
        Marvin N. Wright\textsuperscript{2} 
}
\authorrunning{
David S. Watson \and 
Marvin N. Wright
}

\institute{\textsuperscript{1}Department of Statistical Science \\
              University College London \\
              London, UK\\ 
              \textsuperscript{*}Corresponding Author \email{david.watson@ucl.ac.uk}           
          \and \\
          \textsuperscript{2}Leibniz Institute for Prevention Research and Epidemiology -- BIPS \& \\ 
              Faculty of Mathematics and Computer Science 
              \\ University of Bremen \\
              Bremen, Germany
}



\date{Received: date / Accepted: date}

\maketitle

\begin{abstract}
We propose the conditional predictive impact (CPI), a consistent and unbiased estimator of the association between one or several features and a given outcome, conditional on a reduced feature set. Building on the knockoff framework of \citet{Candes2018}, we develop a novel testing procedure that works in conjunction with any valid knockoff sampler, supervised learning algorithm, and loss function. The CPI can be efficiently computed for high-dimensional data without any sparsity constraints. We demonstrate convergence criteria for the CPI and develop statistical inference procedures for evaluating its magnitude, significance, and precision. These tests aid in feature and model selection, extending traditional frequentist and Bayesian techniques to general supervised learning tasks. The CPI may also be applied in causal discovery to identify underlying multivariate graph structures. We test our method using various algorithms, including linear regression, neural networks, random forests, and support vector machines. Empirical results show that the CPI compares favorably to alternative variable importance measures and other nonparametric tests of conditional independence on a diverse array of real and simulated datasets. Simulations confirm that our inference procedures successfully control Type I error and achieve nominal coverage probability. Our method has been implemented in an \texttt{R} package, \texttt{cpi}, which can be downloaded from \url{https://github.com/dswatson/cpi}. 
\end{abstract}

\section{Introduction}\label{sec:intro}
Variable importance (VI) is a major topic in statistics and machine learning. It is the basis of most if not all feature selection methods, which analysts use to identify key drivers of variation in an outcome of interest and/or create more parsimonious models \citep{Guyon2003,Kuhn2019,Meinshausen2010}. Many importance measures have been proposed in recent years, either for specific algorithms or more general applications. Several different notions of VI -- some overlapping, some inconsistent -- have emerged from this literature. We examine these in greater detail in Sect.~\ref{sec:vim}.

One fundamental difference between various importance measures is whether they test the marginal or conditional independence of features. To evaluate response variable $Y$'s marginal dependence on predictor $X_j$, we test against the following hypothesis:
\begin{center}
    \text{Marginal:}\\ 
    $H_0: X_j \independent Y, \textbf{X}_{-j}$
\end{center}
where $\textbf{X}_{-j}$ denotes a set of covariates.\footnote{We denote variables using uppercase italicized letters, e.g. $X$; matrices using uppercase bold letters, e.g. $\textbf{X}$; scalars using lowercase italicized letters, e.g. $x$; and row-vectors using lowercase bold letters, e.g. $\textbf{x}$.} A measure of conditional dependence, on the other hand, tests against a different null hypothesis:
\begin{center}
    \text{Conditional:}\\ 
    $H_0: X_j \independent Y | \textbf{X}_{-j}$
\end{center}
Note that $X_j$'s marginal VI may be high due to its association with either $Y$ or $\textbf{X}_{-j}$. This is why measures of marginal importance tend to favor correlated predictors. Often, however, our goal is to determine whether $X_j$ adds any \textit{new} information -- in other words, whether $Y$ is dependent on $X_j$ even after conditioning on $\textbf{X}_{-j}$. This becomes especially important when the assumption of feature independence is violated.

Tests of conditional independence (CI) are common in the causal modelling literature. For instance, the popular PC algorithm \citep{Spirtes2000}, which infers a set of underlying directed acyclic graphs (DAGs) consistent with some observational data, relies on the results of CI tests to recursively remove the edges between nodes. Common parametric examples include the partial correlation test for continuous variables or the $\chi^2$ test for categorical data. A growing body of literature in recent years has examined nonparametric alternatives to these options. We provide an overview of several such proposals in Sect.~\ref{sec:cit}.

In this paper, we introduce a new CI test to measure VI. The conditional predictive impact (CPI) quantifies the contribution of one or several features to a given algorithm's predictive performance, conditional on a complementary feature subset. Our work relies on so-called ``knockoff'' variables (formally defined in Sect.~\ref{sec:ko}) to provide negative controls for feature testing. Because knockoffs are, by construction, exchangeable with their observed counterparts and conditionally independent of the response, they enable a paired testing approach without any model refitting. Unlike the original knockoff filter, however, our methods are not limited to certain types of datasets or algorithms.

The CPI is extremely general. It can be used with any combination of knockoff sampler, supervised learner, and loss function. It can be efficiently computed in low or high dimensions without sparsity constraints. We demonstrate that the CPI is an unbiased estimator, provably consistent under minimal assumptions. We develop statistical inference procedures for evaluating its magnitude, precision, and significance. Finally, we demonstrate the measure's utility on a variety of real and simulated datasets.  

The remainder of this paper is structured as follows. We review related work in Sect.~\ref{sec:relatedwork}. We present theoretical results in Sect.~\ref{sec:theory}, where we also outline an efficient algorithm for estimating the CPI, along with corresponding $p$-values and confidence intervals. We test our procedure on real and simulated data in Sect.~\ref{sec:experiments}, comparing its performance with popular alternatives under a variety of regression and classification settings. Following a discussion in Sect.~\ref{sec:discussion}, we conclude in Sect.~\ref{sec:conclusion}.

\section{Related Work}\label{sec:relatedwork} In this section, we survey the relevant literature on VI estimation, CI tests, and the knockoff filter.

\subsection{Variable Importance Measures}\label{sec:vim}
The notion of VI may feel fairly intuitive at first, but closer inspection reveals a number of underlying ambiguities. One important dichotomy is that between global and local measures, which respectively quantify the impact of features on all or particular predictions. This distinction has become especially important with the recent emergence of interpretable machine learning techniques designed to explain individual outputs of black box models \citep[e.g.,][]{Datta2016,Lundberg2017,Ribeiro2016,Wachter2018}. In what follows, we restrict our focus to global importance measures. 

Another important dichotomy is that between model-specific and model-agnostic approaches. For instance, a number of methods have been proposed for estimating importance in linear regression \citep{barber2015,Gromping2007,Lindeman1980}, random forests \citep{Breiman2001,Kursa2010,Strobl2008}, and neural networks \citep{Bach2015,Shrikumar2017,sundar2017}. These measures have the luxury of leveraging an algorithm's underlying assumptions and internal architecture for more precise and efficient VI estimation. 

Other, more general techniques have also been developed. \Citet{VanderLaan2006} derives efficient influence curves and inference procedures for a variety of VI measures. \citet{Hubbard2018} build on this work, proposing a data-adaptive method for estimating the causal influence of variables within the targeted maximum likelihood framework \citep{VanderLaan2018}. \citet{Williamson2017} describe an ANOVA-style decomposition of a regressor's $R^2$ into feature-wise contributions. \citet{Feng2018} design a neural network to efficiently compute this decomposition using multi-task learning. \citet{fisher2019} propose a number of ``reliance'' statistics, calculated by integrating a loss function over the empirical distribution of covariates while holding a given feature vector constant.

Perhaps the most important distinction between various competing notions of VI is the aforementioned split between marginal and conditional measures. The topic has received considerable attention in the random forest literature, where Breiman's popular permutation technique \citep{Breiman2001} has been criticized for failing to properly account for correlations between features \citep{Gregorutti2015,Nicodemus2010}. Conditional alternatives have been developed \citep{Mentch2016,Strobl2008}, but we do not consider them here, as they are specific to tree ensembles. 

Our proposed measure resembles what \citet{fisher2019} call ``algorithm reliance'' (AR). The authors do not have much to say about AR in their paper, the majority of which is instead devoted to two related statistics they term ``model reliance'' (MR) and ``model class reliance'' (MCR). These measure the marginal importance of a feature subset in particular models or groups of models, respectively. Only AR measures the importance of the subset conditional on remaining covariates for a given supervised learner, which is our focus here. \citet{fisher2019} derive probabilistic bounds for MR and MCR, but not AR. They do not develop hypothesis testing procedures for any of their reliance statistics.

\subsection{Conditional Independence Tests}\label{sec:cit}
CI tests are the cornerstone of constraint-based and hybrid methods for causal graph inference and Bayesian network learning \citep{Koller2009,Korb2009,Scutari2014}. Assuming the causal Markov condition and faithfulness -- which together state (roughly) that statistical independence implies graphical independence and vice versa -- a number of algorithms have been developed that use CI tests to discover an equivalence class of DAGs consistent with a set of observational data \citep{Maathuis2009,Verma1990,Spirtes2000}. 

\citet{Shah2018} have shown that there exists no uniformly valid CI test. Parametric assumptions are typically deployed to restrict the range of alternative hypotheses, which is default behavior for most causal discovery software \citep[e.g.,][]{Kalisch2012,Scutari2010}. However, more flexible methods have been introduced. Much of this literature relies on techniques that embed the data in a reproducing kernel Hilbert space (RKHS). For instance, \citet{Fukumizu2008} use a normalized cross-covariance operator to test the association between features in the RKHS. A null distribution is approximated via permutation. \citet{Doran2014} build on \citet{Fukumizu2008}'s work with a modified permutation scheme intended to capture the effects of CI. \citet{Zhang2012} derive a test statistic from the traces of kernel matrices, using a gamma null distribution to compute statistical significance. 

Another general family of methods for CI testing is rooted in information theory. \citet{fleuret2004} proposes a fast binary variable selection procedure using conditional mutual information. Similar techniques have been used to infer directionality in networks \citep{martin2008}, cluster features together for dimensionality reduction \citep{martinez2010}, and reason about the generalization properties of supervised learners \citep{steinke2020}. These approaches typically rely either on discretization procedures to convert all inputs to categorical data, or strong parametric assumptions to handle continuous spaces.

Several authors have proposed alternative tests to avoid the inefficiencies of kernel methods and the binning often required by information theoretic algorithms. For instance, \citet{Strobl2017} employ a fast Fourier transform to reduce the complexity of matrix operations. Methods have been developed for estimating regularized, nonlinear partial correlations \citep{Ramsey2014,Shah2018}. \citet{Lei2018} and \citet{Rinaldo2019} study the leave-one-covariate-out (LOCO) procedure, in which an algorithm is trained on data with and without the variable of interest. The predictive performance of nested models is compared to evaluate the conditional importance of the dropped feature. 

Our proposal is conceptually similar to LOCO, which can in principle be extended to feature subsets of arbitrary dimension. The method enjoys some especially nice statistical properties when used in conjunction with sample splitting. For instance, \citet{Rinaldo2019} derive a central limit theorem for LOCO parameters, while \citet{Lei2018} prove finite sample error control using conformal inference. However, retraining an algorithm for each CI test is potentially infeasible, especially for complex learners and/or large datasets. With knockoffs, we can directly import LOCO's statistical guarantees without any model refitting.

\subsection{The Knockoff Framework}\label{sec:ko}
Our work builds on the knockoff procedure originally conceived by \citet{barber2015} and later refined by \citet{Candes2018}. Central to this approach is the notion of a \textit{knockoff variable}. Given an $n \times p$ input matrix $\textbf{X}$, we define a knockoff matrix of equal dimensionality $\tilde{\textbf{X}}$ as any matrix that meets the following two criteria:

\begin{enumerate}[(a)]
\item \textbf{Pairwise exchangeability.} For any proper subset $S \subset [p] = (1, \dots, p)$:
\begin{align*}
(\textbf{X}, \tilde{\textbf{X}})_{swap(S)} \stackrel{d}{=} (\textbf{X}, \tilde{\textbf{X}})
\end{align*}
where $\stackrel{d}{=}$ represents equality in distribution and the swapping operation is defined below.
\item \textbf{Conditional independence}. $\tilde{\textbf{X}} \independent Y | \textbf{X}$.
\end{enumerate}
A swap is obtained by switching the entries $X_j$ and $\tilde{X}_j$ for each $j \in S$. For example, with $p = 3$ and $S = \{1, 3\}$:
\begin{align*}
    (X_1, X_2, X_3, \tilde{X}_1, \tilde{X}_2, \tilde{X}_3)_{swap(S)} \stackrel{d}{=} (\tilde{X}_1, X_2, \tilde{X}_3, X_1, \tilde{X}_2, X_3)
\end{align*}
Knockoffs provide negative controls for conditional independence testing. The intuition behind the method is that if $X_j$ does not significantly outperform $\tilde{X}_j$ by some relevant importance measure, then the original feature may be safely removed from the final model. 

Practical implementation requires both a method for generating knockoffs and a decision procedure for variable selection. The subject has quickly become a busy one in statistics and machine learning, with most authors focusing on the former task. In this paper, we instead tackle the latter, developing a general framework for testing conditional VI.

Constructing nontrivial knockoffs is a considerable challenge. Numerous methods have been proposed, including but not limited to: second-order Gaussian knockoffs \citep{Candes2018}; conditional permutation sampling \citep{Berrett2018}; hidden Markov model sampling \citep{Sesia2019}; conditional density estimation \citep{Tansey2018}; generative deep neural networks \citep{Romano2018,jordon2019}; and Metropolis-Hastings sampling \citep{Bates2019}. A complete review of these proposals is beyond the scope of this paper. \citet{Bates2019} demonstrate that no efficient knockoff sampler exists for arbitrary probability distributions, suggesting that algorithms will have to make some assumptions about the data generating process to strike a reasonable balance between sensitivity and specificity.  

The original knockoff papers introduce a novel algorithm for controlling the false discovery rate (FDR) in variable selection problems. The goal is to find the minimal subset $\mathcal{S} \subset [p]$ such that, conditional on $\{X_j\}_{j \in \mathcal{S}}$, $Y$ is independent of all other variables. Call this the Markov blanket of $Y$ \citep{Pearl1988}. Null features form a complementary set $\mathcal{R} = [p] \setminus \mathcal{S}$ such that $k \in \mathcal{R}$ if and only if $X_k \independent Y | \{X_j\}_{j \in \mathcal{S}}$. The FDR is given by the expected proportion of false positives among all declared positives:
\begin{align*}
    \text{FDR} = \mathds{E}\left[\frac{\lvert{\widehat{\mathcal{S}}} \cap \mathcal{R} \rvert}{\lvert{\widehat{\mathcal{S}}}\rvert \vee 1}\right]
\end{align*}
where $\widehat{\mathcal{S}}$ is the output of the decision procedure and the ``$\vee$ 1'' in the denominator enforces the convention that $\text{FDR}= 0$ when $\lvert{\widehat{\mathcal{S}}}\rvert = 0$.

\citet{barber2015} demonstrate a method for guaranteeing finite sample FDR control when (i) statistics for null variables are symmetric about zero and (ii) large positive statistics indicate strong evidence against the null. We will henceforth refer to this method as the adaptive thresholding test (ATT). Unlike other common techniques for controlling the FDR \citep[e.g.,][]{Benjamini1995,benjamini2001,Storey2002}, the ATT does not require $p$-values as an intermediate step. \citet{Candes2018} argue that this is a benefit in high-dimensional settings, where $p$-value calculations can be unreliable. 

Acknowledging that $p$-values may still be desired in some applications, however, Cand\`{e}s et al. also propose the conditional randomization test (CRT), which provides one-sided Monte Carlo $p$-values by repeatedly sampling from the knockoff distribution. Experiments indicate that the CRT is slightly more powerful than the ATT, but the authors caution that the former is very computationally intensive and do not recommend it for large datasets. That has not stopped other groups from advancing formally similar proposals \citep[e.g.,][]{Berrett2018,Tansey2018}.

We highlight several important shortcomings of the ATT:
\begin{enumerate}
\itemsep-.25em
    \item Not all algorithms provide feature scoring statistics.
    \item The ATT requires a large number of variables to reliably detect true positives. 
    \item Because the ATT does not perform individual hypothesis tests, it cannot provide confidence or credible intervals for particular variables. 
\end{enumerate}
In what follows, we present alternative inference procedures for conditional independence testing designed to address all three issues. 

\section{Conditional Predictive Impact}\label{sec:theory}
The basic intuition behind our approach is that important features should be \textit{informative} --- that is, their inclusion should improve the predictive performance of an appropriate algorithm as measured by some preselected loss function. Moreover, the significance of improvement should be quantifiable so that error rates can be controlled at user-specified levels.

Consider an $n \times p$ feature matrix $\textbf{X} \in \mathcal{X}$ and corresponding $n \times 1$ response variable $Y \in \mathcal{Y}$, which combine to form the dataset $\textbf{Z} = (\textbf{X}, Y) \in \mathcal{Z}$.\footnote{We leave the more general case of multidimensional response variables to future work.} Each observation $\textbf{z}_i = (\textbf{x}_i, y_i)$ is an i.i.d. sample from a fixed but unknown joint probability distribution, $\mathds{P}(\textbf{Z}) = \mathds{P}(\textbf{X}, Y)$. Let $\textbf{X}^S \subseteq (X_1, \dots, X_p)$ denote some subset of the feature space whose predictive impact we intend to quantify, conditional on the (possibly empty) set of remaining covariates $\textbf{X}^R = \textbf{X}\setminus\textbf{X}^S$. Data can now be expressed as a triple, $\textbf{Z} = (\textbf{X}^S, \textbf{X}^R, Y)$. We remove the predictive information in $\textbf{X}^S$ while preserving the covariance structure of the predictors by replacing the submatrix with the corresponding knockoff variables, $\tilde{\textbf{X}}^S$, rendering a new dataset, $\tilde{\textbf{Z}} = (\tilde{\textbf{X}}^S,\textbf{X}^R,Y)$. 

Define a function $f \in \mathcal{F}, \mathcal{F}: \mathcal{X} \rightarrow \mathcal{Y}$ as a mapping from features to outcomes. We evaluate a model's performance using some real-valued, nonnegative loss function $L: \mathcal{F} \times \mathcal{Z} \rightarrow \mathds{R}_{\geq 0}$. Define the risk of $f$ with respect to \textbf{Z} as its expected loss over the joint probability distribution $\mathds{P}(\textbf{Z})$: 
\begin{align*} 
R(f, \textbf{Z}) = \mathds{E}[L(f,\textbf{Z})]
\end{align*}
Our strategy is to replace the conditional null hypothesis defined in Sect.~\ref{sec:intro} with the following:
\begin{align*}
    \text{Conditional Predictive:}\\ 
    H_0: R(f, \textbf{Z}) \geq R(f, \tilde{\textbf{Z}})
\end{align*}
In other words, we test whether the model performs better using the original or the knockoff data. We note that this is a potentially weaker $H_0$ than that of true conditional independence (Sect.~\ref{sec:intro}), as many popular loss functions restrict attention to just the first moment. However, we argue that if such a loss function is appropriate, then a conditional predictive test is a better choice than a conditional independence test. For instance, if the goal is simply to minimize out-of-sample $L_2$ loss, then we have no need for features that do not improve mean square error, even if they encode information about higher moments (e.g., predictive variance or skewness). Alternatively, if confidence intervals are important for a given task, then the loss function should reflect that, as likelihood-based measures do. In general, recovering the full Markov blanket of $Y$ may be unnecessary.

\subsection{Consistency and Convergence}\label{sec:cc}
The CPI of submatrix $\textbf{X}^S$ measures the extent to which the feature subset improves predictions made using model $f$. Assume that the loss function $L$ can be evaluated for each sample $i$.\footnote{For loss functions that do not have this property, such as the area under the receiver operating characteristic curve, the following arguments can easily be modified to apply to each fold in a cross-validation.} We define the following random variable:
\begin{equation}
    \Delta_i = L(f, \tilde{\textbf{z}}_i) - L(f, \textbf{z}_i)
\end{equation}
This vector represents the difference in sample-wise loss between predictions made using knockoff data and original data. We define the CPI by taking its expectation:
\begin{equation}
    \text{CPI}(\textbf{X}^S) = \mathds{E}[\Delta]
\end{equation}
Note that the CPI is always a function of some feature subset $\textbf{X}^S$. We suppress the dependency for notational convenience moving forward. 

To consistently estimate this statistic, it is necessary and sufficient to show that we can consistently estimate the risk of model $f$. The population parameter $R(f, \textbf{Z})$ is estimated using the empirical risk formula: 
\begin{equation} \label{eq:risk}
    R_\text{emp}(f, \textbf{Z}) = \frac{1}{m} \sum_{i=1}^{m} L(f,\textbf{z}_i)
\end{equation}
Our goal in estimating risk is to evaluate how well the model generalizes beyond its training data, so the $m$ samples in Eq. \ref{eq:risk} constitute a test set drawn independently from $\textbf{Z}$, distinct from the $n$ samples used to fit $f$. In practice, this is typically achieved by a resampling procedure such as cross-validation or bootstrapping. In what follows, we presume that unit-level loss $L(f, \textbf{z}_i)$ is always an out-of-sample evaluation, such that $f$ was trained on data excluding $\textbf{z}_i$. 

The empirical risk minimization (ERM) principle is a simple decision procedure in which we select the function $f$ that minimizes empirical risk in some function space $\mathcal{F}$. A celebrated result of \citet{Vapnik1971}, independently derived by \citet{Sauer1972} and \citet{Shelah1972}, is that the ERM principle is consistent with respect to $\mathcal{F}$ if and only if the function space is of finite VC dimension. Thus, for any algorithm that meets this criterion, the empirical risk $R_\text{emp}(f, \textbf{Z})$ converges uniformly in probability to $R(f, \textbf{Z})$ as $n \rightarrow \infty$, which means the estimate
\begin{equation}
\begin{aligned}
    \widehat{\text{CPI}} &={} \frac{1}{n} \sum_{i=1}^{n} L(f, \tilde{\textbf{z}}_i) - L(f, \textbf{z}_i) \\ 
    &= R_\text{emp}(f, \tilde{\textbf{Z}}) - R_\text{emp}(f, \textbf{Z}) \\
    &= \frac{1}{n} \sum_{i=1}^{n} \Delta_{i}
\end{aligned}
\end{equation}
is likewise guaranteed to converge. Because the CPI inherits the convergence properties of the learner $f$, it imposes no additional smoothness, sparsity, parametric, or dimensionality constraints upon the data. 

Though finite complexity thresholds have been derived for many algorithms -- e.g., projective planes, decision trees, boosting machines, and neural networks \citep{Shalev-Shwartz2014} -- it is worth noting that some popular supervised learners do in fact have infinite VC dimension. This is the case, for instance, with methods that rely on the radial basis function kernel, widely used in support vector machines and Gaussian process regression. The learning theoretic properties of these algorithms are better described with other measures such as the Rademacher complexity and PAC-Bayes bounds \citep{guedj2019primer}. However, as we show in Sect.~\ref{sec:experiments}, the CPI shows good convergence properties even when used with learners of infinite VC dimension.

Inference procedures for the CPI can be designed using any paired difference test. Familiar frequentist examples include the $t$-test and the Fisher exact test, which we use for large- and small-sample settings, respectively. Bayesian analogues can easily be implemented as well. \citet{Rouder2009} advocate an analytic strategy for calculating Bayes factors for $t$-tests. \citet{Wetzels2009} and \citet{Kruschke2013} propose more general methods based on Markov chain Monte Carlo sampling, although they differ in their proposed priors and decision procedures. Care should be taken when selecting a prior distribution in the Bayesian setting, especially with small sample sizes. Tools for Bayesian inference are implemented in the \texttt{cpi} package; however, for brevity's sake, we restrict the following sections to frequentist methods.

\subsection{Large Sample Inference: Paired $t$-tests}\label{sec:ttest}
By the central limit theorem, empirical risk estimates for functions of finite VC dimension will tend to be normally distributed around the true population parameter value. We therefore use paired, one-sided $t$-tests to evaluate statistical significance when samples are sufficiently large. As $n$ grows, these results will converge on $z$-test results, although the latter procedure technically presumes known variance of $\Delta$. Since we do not have \emph{a priori} access to this parameter, we opt instead for $t$-tests, which are roughly equivalent for $n \geq 30$.

The variable $\Delta$ has mean $\widehat{\text{CPI}}$ and standard error $\text{SE} = s/\sqrt{n}$, where
\begin{align*}
    s = \sqrt{\frac{1}{n-1}\sum_{i=1}^{n} (\Delta_i - \widehat{\text{CPI}})^2}.
\end{align*}
The $t$-score for $\widehat{\text{CPI}}$ is given by $t = \widehat{\text{CPI}}/\text{SE}$, and we compute $p$-values by comparing this statistic to the most tolerant distribution consistent with $H_0: R(f, \textbf{Z}) \geq R(f,\tilde{\textbf{Z}})$, namely $t_{n-1}$. To control Type I error at level $\alpha$, we reject $H_0$ for all $t$ greater than or equal to the $(1-\alpha)$ quantile of $t_{n-1}$. This procedure can easily be modified to adjust for multiple testing. 

We can relax the assumption of homoskedasticity if reliable estimates of predictive precision are available. Construct a $2n\times(n+1)$ feature matrix $\textbf{X}$ with columns for each unit $i = \{1,\dots,n\}$, as well as an indicator variable for data type $D$ (original vs. knockoff). Let \textbf{W} be a $2n\times2n$ diagonal matrix such that $\textbf{W}_{ii}$ denotes the weight assigned to the $i$th prediction. For instance, in a regression setting, this could be the inverse of the expected residual variance for $i$.  Then solve a weighted least squares regression, with the response variable $\textbf{y}$ equal to the observed loss for each unit-data type combination: 
\begin{align*}
    \bm{\hat{\gamma}} = (\textbf{X}^\top\textbf{WX})^{-1}\textbf{X}^\top\textbf{Wy}
\end{align*}
The $t$-statistic and $p$-value associated with coefficient $\hat{\gamma}_D$ can then be used to test the CPI of the substituted variable(s) under a heteroskedastic error model. 

Confidence intervals around $\widehat{\text{CPI}}$ may be constructed in the typical manner. The lower bound is set by subtracting from our point estimate the product of SE and $F^{-1}_{n-1} (1-\alpha)$, where $F_{n-1}(\cdot)$ denotes the CDF of $t_{n-1}$. Using this formula, we obtain a 95\% confidence interval for $\widehat{\text{CPI}}$ by calculating $[\widehat{\text{CPI}} - \text{SE} \times F^{-1}_{n-1}(0.95), \infty$). As $n$ grows large, this interval converges to the Wald-type interval, $[\widehat{\text{CPI}} - \text{SE} \times \Phi^{-1}(0.95), \infty)$, where $\Phi$ represents the standard normal CDF. 

The $t$-testing framework also allows for analytic power calculations. Let $t^*$ denote the critical value $t^* = F^{-1}_{n-1}(1-\alpha)$. Then Type II error is given by the formula $\beta = F_{n-1}(t^* - \delta)$, where $\delta$ represents the postulated effect size. Statistical power is just the complement $1-\beta$, and rearranging this equation with simple algebra allows us to determine the sample size required to detect a given effect at some fixed Type I error $\alpha$.

\subsection{Small Sample Inference: Fisher Exact Tests}\label{sec:fishertest}
The applicability of the central limit theorem is dubious when sample sizes are small. In such cases, exact $p$-values may be computed for a slightly modified null hypothesis using Fisher's method \citep{Fisher1935}. Rather than focusing on overall risk, this null hypothesis states that replacing $\textbf{X}^S$ with the knockoff submatrix $\tilde{\textbf{X}}^S$ has no impact on unit-level loss. More formally, we test against the following:
\begin{align*}
    H_0^{\text{FEP}}: L(f, \textbf{z}_i) \geq L(f, \tilde{\textbf{z}}_i),\quad i = 1, \dots, n.
\end{align*}
Under this null hypothesis, which is sufficient but unnecessary for the conditional predictive $H_0$, we may implement a permutation scheme in which the CPI is calculated for all possible assignments of data type $D$. Consider a $2n \times 3$ matrix with columns for unit index $U = \{1, 1, \dots, n, n\}$, data type $D \in \{0, 1\}$, and loss $L$. We permute the rows of $D$ subject to the constraint that every sample's loss is recorded under both original and knockoff predictions. For each possible vector $D$, compute the resulting CPI and compare the value of our observed statistic, $\widehat{\text{CPI}}$, to the complete distribution. Note that this paired setup dramatically diminishes the possible assignment space from an unmanageable $( \genfrac{}{}{0pt}{}{2n}{n} )$, corresponding to a Bernoulli trial design, to a more reasonable $2^n$. The one-tailed Fisher exact $p$-value (FEP) is given by the formula:
\begin{align*}
    \text{FEP}(\widehat{\text{CPI}}) = \frac{1}{2^n} \sum_{b=1}^{2^n} \mathds{1}[{\widetilde{\text{CPI}}_b \geq \widehat{\text{CPI}}}]
\end{align*}
where $\mathds{1}[\cdot]$ represents the indicator function and $\widetilde{\text{CPI}}_b$ is the CPI resulting from the $b^\text{th}$ permutation of $D$.
	
To construct a confidence interval for $\widehat{\text{CPI}}$ at level $1 - \alpha$, we use our empirical null distribution. Find the critical value $\text{CPI}^*$ such that $\text{FEP}(\text{CPI}^*) = \alpha$. Then a $(1 - \alpha) \times 100\%$ confidence interval for $\widehat{\text{CPI}}$ is given by $[\widehat{\text{CPI}} - \text{CPI}^*, \infty)$. For $n$ large, approximate calculations can be made by sampling from the set of $2^n$ permissible permutations. In this case, however, it is important to add 1 to both the numerator and denominator to ensure unbiased inference \citep{Phipson2010}.

\subsection{Computational Complexity}
To summarize, we outline our proposed algorithm for testing the conditional importance of feature subsets for supervised learners in pseudocode below. 

\vspace*{5mm}
\begin{algorithm}[H]
\caption{CPI Algorithm}
\KwIn{Dataset $\textbf{Z}$, submatrix $\textbf{X}^S$, supervised learner $a$, risk functional $R$, knockoff sampler $g$, risk estimator $k$, inference procedure $h$}
\begin{enumerate}
\itemsep-.25em
    \item Train $a$ on $\textbf{Z}$ to create $f$
    \item Apply $g$ to generate the knockoff matrix $\tilde{\textbf{X}}^S$
    \item Use risk estimator $k$ to compute each $L(f, \textbf{z}_i)$ and $L(f, \widetilde{\textbf{z}}_i)$
    \item Compute $\widehat{\text{CPI}} = n^{-1}\sum_{i=1}^n L(f, \widetilde{\textbf{z}}_i) - L(f, \textbf{z}_i)$ 
    \item Apply inference procedure $h$ to determine associated $p$-value ($p$) and confidence interval (ci)
\end{enumerate}
\KwOut{$\widehat{\text{CPI}}$, $p$, ci}
\end{algorithm}
\vspace*{5mm}

This algorithm executes in $\mathcal{O}(ak + g + h)$ time and $\mathcal{O}(a + k + g + h)$ space. We take the complexity of the learner $a$ and knockoff sampler $g$ to be given. The empirical risk estimator $k$ can be made more or less complex depending on the resampling procedure. The most efficient option for evaluating generalization error is the holdout method, in which a model is trained on a random subset of the available data and tested on the remainder. Unfortunately, this procedure can be unreliable with small sample sizes. Popular alternatives include the bootstrap and cross-validation. Both require considerable model refitting, which can be costly when $a$ is complex. 

The inference procedure $h$ is quite efficient in the parametric case -- on the order of $\mathcal{O}(n)$ for the $t$-test -- but scales exponentially with the sample size when using the permutation-based approach. As noted above, the time complexity of the Fisher test can be bounded by setting an upper limit on the number of permutations $B$ used to approximate the empirical null distribution. The standard error of a $p$-value estimate made using such an approximation is $\sqrt{p^*(1-p^*)/B}$, where $p^*$ represents the true $p$-value. This expression is maximized at $p^* = 0.5$, corresponding to a standard error of $1/(2\sqrt{B})$. Thus, to guarantee a standard error of at most $0.001$, it would suffice to use $B = 250,000$ permutations, an eminently feasible computation using parallel processors. Space complexity is dominated by the learner $a$ and knockoff sampler $g$ because most resampling procedures refit the same learner a fixed number of times and the inference procedures described above execute in $\mathcal{O}(1)$ space.

\section{Experiments}\label{sec:experiments}
All experiments were conducted in the \texttt{R} statistical computing environment, version 3.5.1. Code for reproducing all results and figures can be found in our dedicated GitHub repository: \url{https://github.com/dswatson/cpi_paper}.

\subsection{Simulated Data}\label{sec:sim}
We report results from a number of simulation studies. First, we analyze the statistical properties of our proposed tests under null and alternative hypotheses. We proceed to compare the sensitivity and specificity of the CPI to those of several alternative measures. 

Data were simulated under four scenarios, corresponding to all combinations of independent vs. correlated predictors and linear vs. nonlinear outcomes. Because conditional importance is most relevant in the case of correlated predictors, results for the two scenarios with independent features are left to the supplement. In the linear setting, variables were drawn from a multivariate Gaussian distribution $\mathcal{N}(0,\Sigma)$, with covariance matrix $\Sigma_{ij} = 0.5^{|i-j|}$. We vary dimensionality $p$ within the range $\{10, 20, 50\}$ and compute the continuous outcome $Y = \textbf{X} \bm{\beta} + \epsilon$, where $\epsilon \sim \mathcal{N}(0,1)$ and $\bm{\beta} = (0.0, 0.1, \dots, 0.9)^\top$ repeated $p/10$ times. In the nonlinear scenario, we keep the same predictors but generate the response from a transformed matrix, $Y = \textbf{X}^* \bm{\beta} + \epsilon$, where 
\begin{align*}
    x_{ij}^* = 
    \begin{cases}
        +1, & \text{if } \Phi^{-1}(0.25) \leq x_{ij} \leq \Phi^{-1}(0.75) \\
        -1, & \text{else}
    \end{cases}
\end{align*}
with the same $\bm{\beta}$ and $\epsilon$ as in the linear case. A similar simulation was performed for a classification outcome, where the response $Y$ was drawn from a binomial distribution with probability $[1 + \exp{(-\textbf{X} \bm{\beta})}]^{-1}$ and $[1 + \exp{(-\textbf{X}^* \bm{\beta})}]^{-1}$ for the linear and nonlinear scenarios, respectively.

Knockoffs for all simulated data were generated using the second-order Gaussian technique described in \citep{Candes2018} and implemented in the \texttt{knockoff} package, version 0.3.2 \citep{knockoff}. 

\subsubsection{Type I and Type II Errors}\label{sec:err}
We simulate $10,000$ datasets with $n=1,000$ observations and compute the CPI using four different learning algorithms: linear/logistic regression (LM), random forests (RF), artificial neural network (ANN), and support vector machine (SVM). Risk was estimated using holdout sampling with a train/test ratio of 2:1. For regression models, we used mean square error (MSE) and mean absolute error (MAE) loss functions; for classification, we used cross entropy (CE) and mean misclassification error (MMCE). We computed $p$-values via the inference procedures described in Sect.~\ref{sec:theory}, i.e. paired $t$-tests and Fisher exact tests. For Fisher tests we used $10,000$ permutations.

Linear and logistic regressions were built using the functions \texttt{lm()} and \texttt{glm()}, respectively, from the \texttt{R} package \texttt{stats} \citep{R2018}. RFs were built using the \texttt{ranger} package \citep{Wright2017}, with $500$ trees. ANNs were built with the \texttt{nnet} package \citep{Venables2002}, with $20$ hidden units and a weight decay of $0.1$. SVMs were built with the \texttt{e1071} package \citep{Meyer2018}, using a Gaussian radial basis function (RBF) kernel and $\sigma = 1$. Unless stated otherwise, all parameters were left to their default values. Resampling was performed with the \texttt{mlr} package \citep{Bischl2016}.

\begin{figure}[p]
\begin{center}
\includegraphics[width=.9\textwidth]{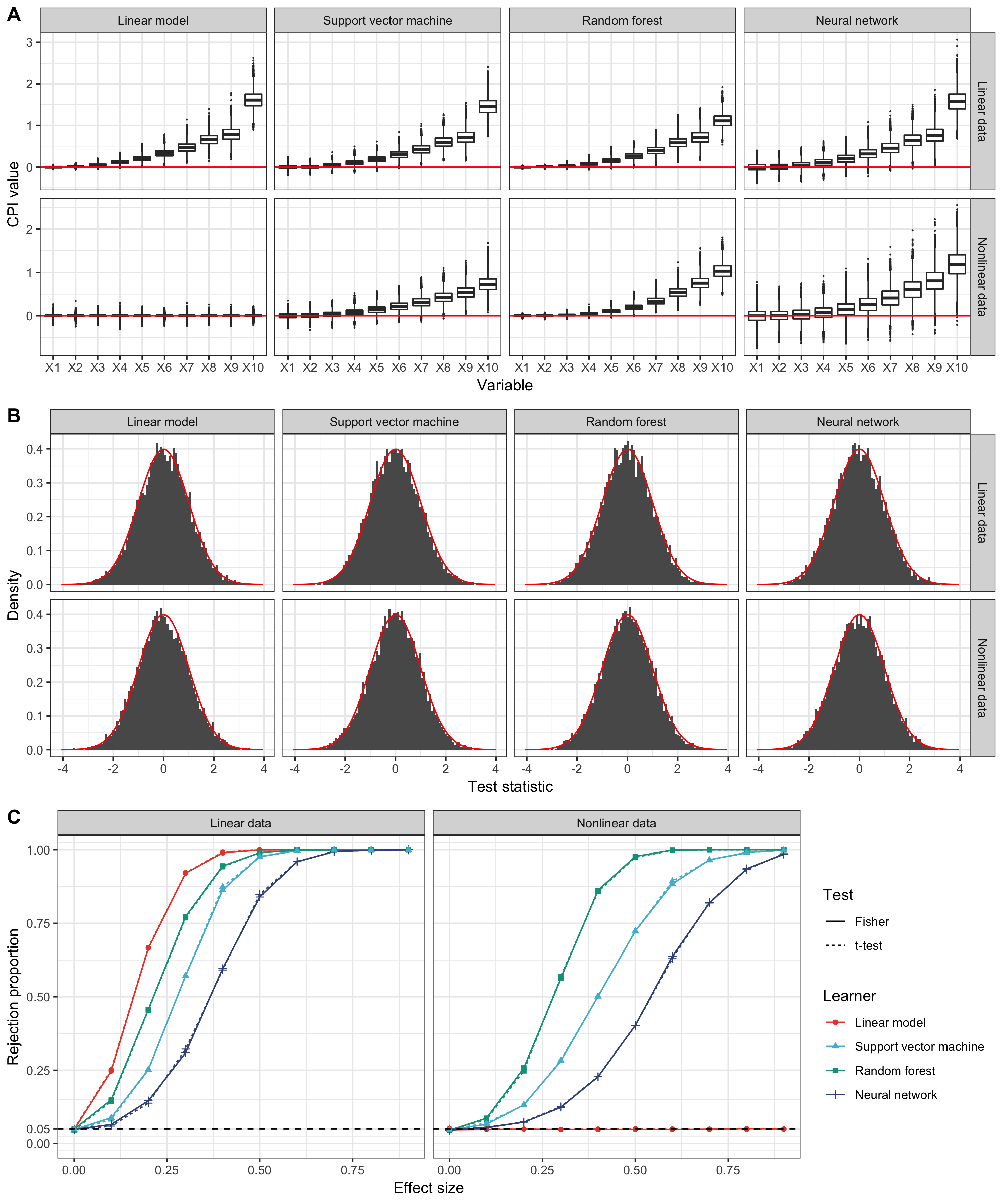}
\end{center}
\vspace*{-5mm}
\caption{Simulation results for continuous outcome with MSE loss and correlated predictors. \textbf{A}: Boxplots of simulated CPI values of variables $X_1, \dots, X_{10}$ with increasing effect size. The red line indicates a CPI value of 0, corresponding to a completely uninformative predictor $X_j$. \textbf{B}: Histograms of simulation replications of $t$-statistics of variables with effect size $0$. The distribution of the expected $t$-statistic under the null hypothesis is shown in red. \textbf{C}: Proportion of rejected hypotheses at $\alpha = 0.05$ as a function of effect size. Results at effect size $0$ correspond to the Type I error, at effect sizes $>0$ to statistical power. The dashed line indicates the nominal level of $\alpha = 0.05$.}
\label{fig:cv-regr-logratio-mse}
\end{figure}

Significance levels for all tests were fixed at $\alpha = 0.05$. For each simulation, we calculated the CPI values, Type I errors, Type II errors, empirical coverage, and $t$-statistics, where applicable. Results for MSE loss with $p = 10$ are shown in Fig.~\ref{fig:cv-regr-logratio-mse}. Similar plots for other loss functions and dimensionalities are presented in Figs.~S1-S10 of the supplement. Coverage probabilities are shown in Tables~S1-S8 of the supplement.

For continuous outcomes, CPI controlled Type I error with all four learners and reached 100\% power under all settings, with the exception of the LM on nonlinear data. We observed no difference between the MSE and MAE loss functions.

We found similar results for categorical outcomes. The CPI controlled Type I error for the MMCE and CE loss functions with all four learners. The LM once again performed poorly on nonlinear data, as expected. The Fisher test had slightly increased power compared to the $t$-test. Statistical power was generally greater with CE loss than with MMCE loss.

\subsubsection{Comparative Performance}\label{sec:compare}
We use the same simulation setup to compare the CPI's performance to that of three other global, nonparametric, model-agnostic measures of CI, each of which relies on identical or stronger testing assumptions:
\begin{description}
\itemsep-.25em
  \item{$\bullet$ \textbf{ANOVA}}: \citet{Williamson2017}'s nonparametric ANOVA-inspired VI.
  \item{$\bullet$ \textbf{LOCO}}: \citet{Lei2018}'s leave-one-covariate-out procedure.
  \item{$\bullet$ \textbf{GCM}}: \citet{Shah2018}'s generalized covariance measure.
\end{description}
Unfortunately, software for \citet{Hubbard2018}'s targeted maximum likelihood VI statistic was still under development at the time of testing, and beta versions generated errors. Cand\`{e}s et al.'s probabilistic knockoff procedure \citep{Candes2018} can be extended to nonparametric models, but requires an algorithm-specific VI measure, which not all learners provide. We consider this method separately in Sect.~\ref{sec:ko_vs_cpi}. Kernel methods do not work with arbitrary algorithms and were therefore excluded. Information theoretic measures typically require a discretization step that does not extend well to high dimensions, and were therefore also excluded. We restrict this section to the regression setting, as none of the other methods considered here are designed for classification problems.

Training and test sets are of equal size, with $n = \{100, 500, 1000\}$. In each case, we fit LM, RF, ANN, and SVM regressions, as described previously. We estimate the VI of all features on the test set for every model. This procedure was repeated 10,000 times. Results for $n = 1,000$ and $p = 10$ are plotted in Fig.~\ref{fig:bigsim}. Similar results for $n = \{100, 500\}$ and $p = \{20, 50, 100\}$ are included in the supplement.

\begin{figure}[!h]
\begin{center}
\includegraphics[width=\textwidth]{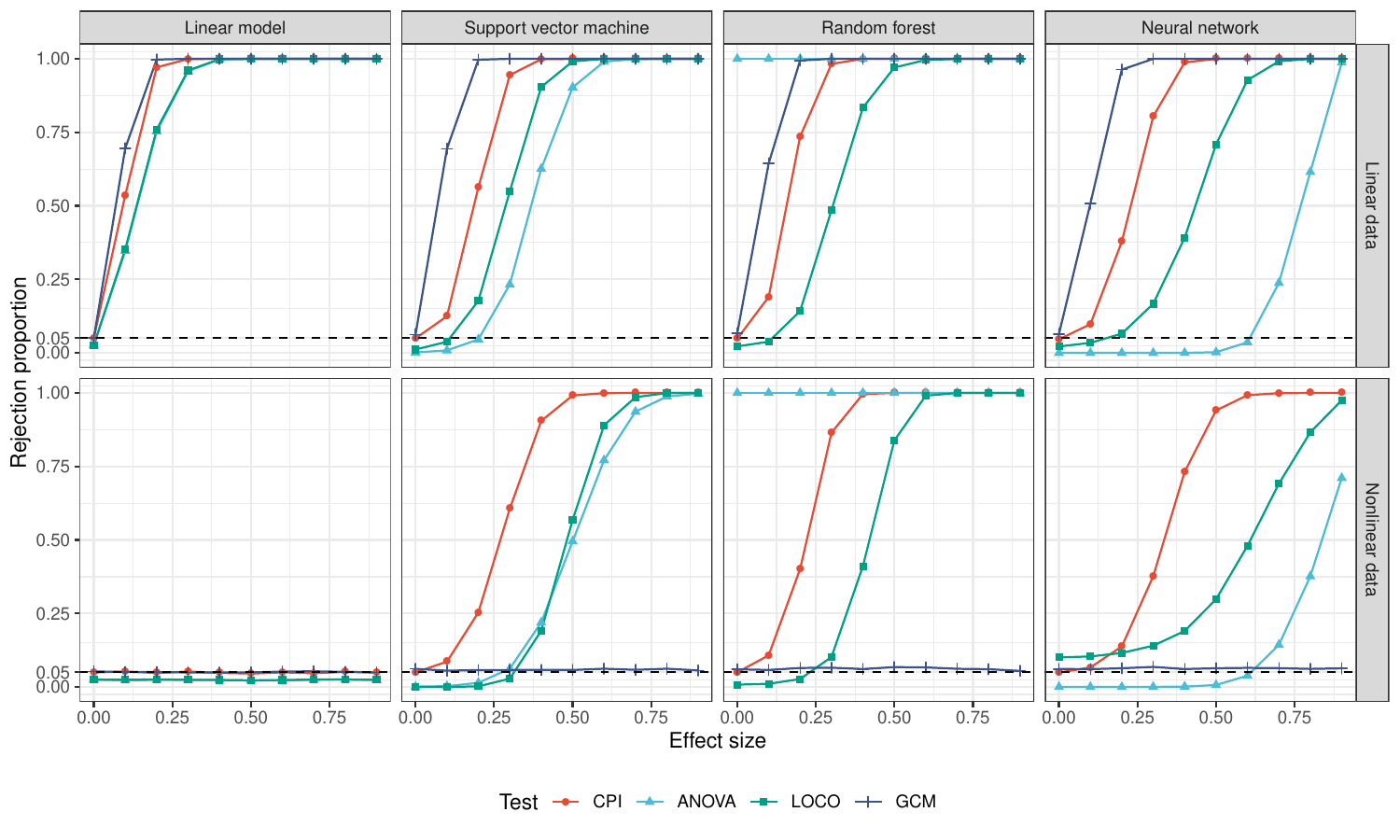}
\end{center}
\caption{Comparative performance of VI measures across different simulations and algorithms. Plots depict the proportion of rejected hypotheses at $\alpha = 0.05$ as a function of effect size. Results at effect size $0$ correspond to Type I error, at effect sizes $>0$ to statistical power. The dashed line indicates the nominal level of $\alpha = 0.05$. These results were computed using training and test samples of $n = 1,000$ and $p = 10$. Similar results were obtained for sample sizes of $n = \{100, 500\}$ and $p = \{20, 50, 100\}$ (see Supplementary Materials).}
\label{fig:bigsim}
\end{figure}

All methods have high Type II error rates when fitting an LM to nonlinear data, highlighting the dangers of model misspecification. GCM appears to dominate in the linear setting, but struggles to detect VI in nonlinear simulations. LOCO is somewhat conservative, often falling short of the nominal Type I error rate under the null hypothesis. However, the method fails to control Type I error in the case of an ANN trained on nonlinear data. The nonparametric ANOVA generally performs poorly, especially with RF regressions, where we observed Type I error rates of up to 100\%.

The CPI outperforms all competitors with nonlinear data, and achieves greater power than ANOVA or LOCO in the linear case. GCM is the only other method to control Type I error under all simulation settings, but it has nearly zero power with nonlinear data. 

\subsubsection{Knockoff Filter}\label{sec:ko_vs_cpi}

To compare the performance of the CPI with that of the original knockoff filter, we followed the simulation procedure described in Sect. 4 of \citep{Candes2018}. A $n = 300 \times p = 1,000$ feature matrix was sampled from a multivariate Gaussian distribution $\mathcal{N}(0,\Sigma)$ with covariance matrix $\Sigma_{ij} = \rho^{|i-j|}$. A continuous outcome $Y$ was calculated as $\textbf{X} \bm{\beta} + \epsilon$, where $\epsilon \sim \mathcal{N}(0,1)$ and the coefficient vector $\bm{\beta}$ contains just 60 nonzero entries, with random signs and variable effect sizes. We vary $\rho$ with fixed nonzero $|\beta| = 1$, and vary effect size with fixed $\rho = 0$. 

We train a series of lasso regressions \citep{Tibshirani1996} on the data using the original design matrix and $10$-fold cross-validation to calculate the CPI, and the expanded $n \times 2p$ design matrix for the knockoff filter. VI for the latter was estimated using the difference statistic originally proposed by \citet{barber2015}: 
\begin{align*}
    W_j = |\hat{\beta}_j| - |\hat{\beta}_{j+p}|
\end{align*}
where $\hat{\beta}_j$ and $\hat{\beta}_{j+p}$ represent the coefficients associated with a feature and its knockoff, respectively, at some fixed value of the Lagrange multiplier $\lambda$. Variables are selected based on the ATT method described in Sect.~\ref{sec:ko}. We tune $\lambda$ via 10-fold cross-validation, per the default settings of the \texttt{glmnet} package \citep{Friedman2010}. Power and FDR are averaged over $10,000$ iterations for each combination of effect size and autocorrelation coefficient. FDR is estimated via the ATT procedure for the knockoff filter and via the Benjamini-Hochberg \citet{Benjamini1995} algorithm for the CPI. 

\begin{figure}[!h]
\begin{center}
\includegraphics[width=\textwidth]{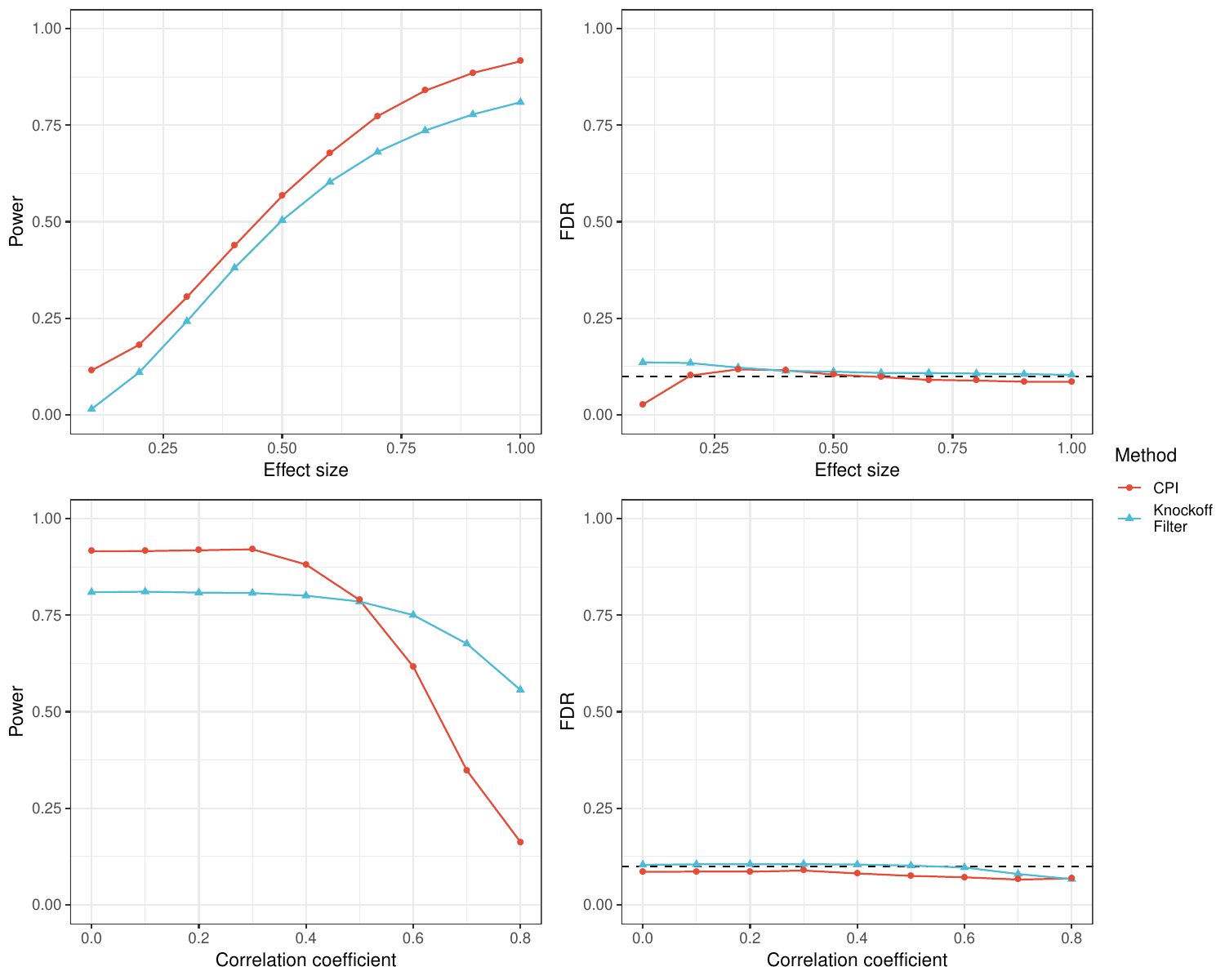}
\end{center}
\caption{Power and FDR as a function of effect size and autocorrelation for CPI and knockoff filter. Target FDR is 10\%. Results are from a lasso regression with $n = 300$ and $p = 1,000$. Each point represents 10,000 replications. Similar results were obtained for $p = 2000$ (see Supplementary Materials).}
\label{fig:kovscpi}
\end{figure}

The CPI is more powerful than the original knockoff filter for all effect sizes at $\rho = 0$, but less powerful for high autocorrelation of $\rho > 0.5$ (see Fig.~\ref{fig:kovscpi}). Both methods generally control the FDR at the target rate of 10\%. The only exceptions are under small effect sizes, where the knockoff filter shows slightly inflated errors. Similar results for $p = 2000$ are included in the supplement.

Note that in addition to being a more powerful test under most conditions, the CPI has other important advantages over the ATT. Whereas the latter can only be applied to algorithms with inbuilt feature scoring statistics, the former requires nothing more than a valid loss function. Whereas the ATT struggles to select important variables in low-dimensional settings, the CPI imposes no dimensionality restraints. Finally, the CPI is more informative, insomuch as it provides feature-level $p$-values and confidence (or credible) intervals.

\subsection{Real Data}\label{sec:real}
In this section, we apply the CPI to real datasets of low- and high-dimensionality. 

\begin{figure}[!h]
\begin{center}
\includegraphics[width=.6\textwidth]{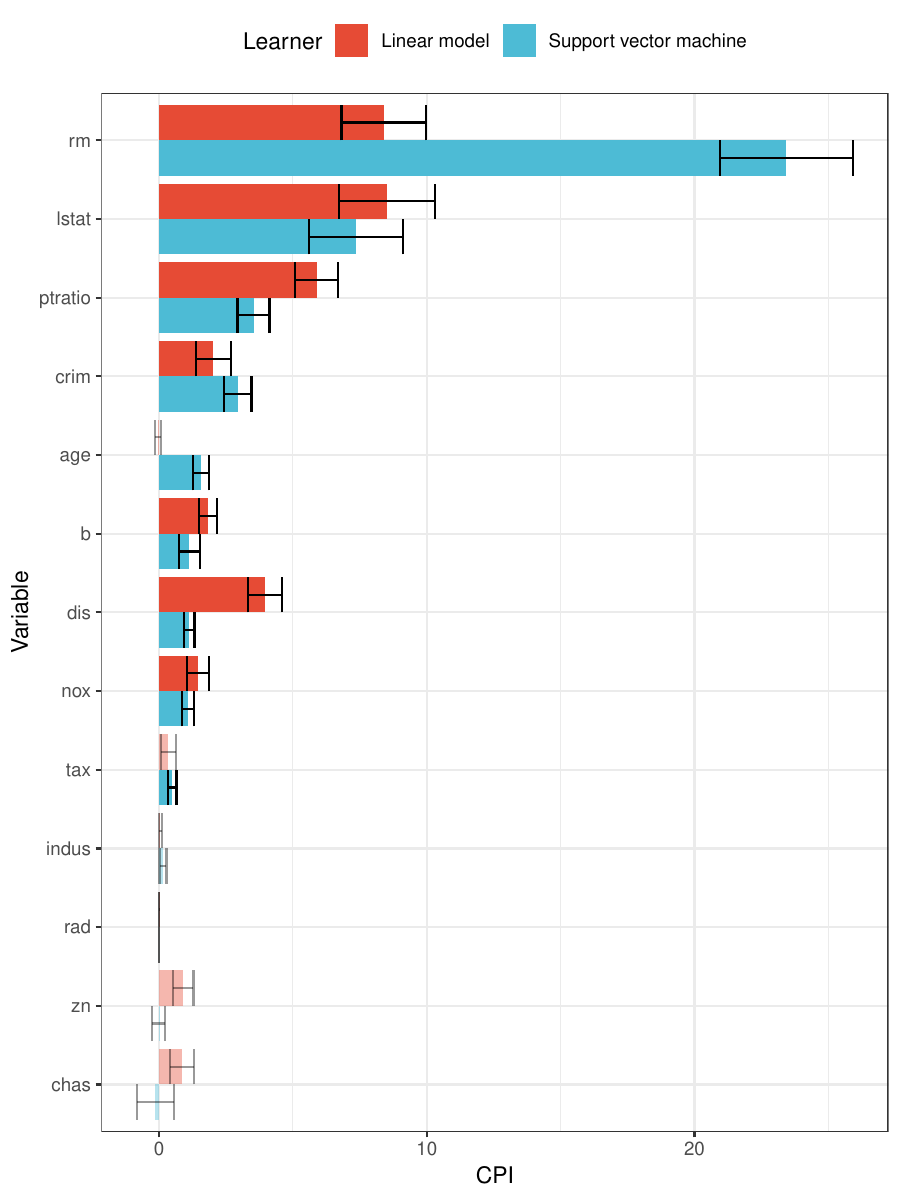}
\end{center}
\caption{Results of the Boston housing experiment. For each variable in the data set, the CPI value is shown, computed with a linear model and a support vector machine. Whiskers represent standard errors. Non-significant variables at $\alpha = 0.05$ after adjustment for multiple testing are shaded. }
\label{fig:boston}
\end{figure}

\subsubsection{Boston Housing}\label{sec:boston}
We analyzed the Boston housing data \citep{Harrison1978}, which consists of 506 observations and 14 variables. This benchmark dataset is available in the UCI Machine Learning Repository \citep{Dua2017}. The dependent variable is the median price of owner-occupied houses in census tracts in the Boston metropolitan area in 1970. The independent variables include the average number of rooms, crime rates, and air pollution. 

Using LM and SVM regressions, we computed CPI, standard errors, and $t$-test $p$-values for each feature, adjusting for multiple testing using Holm's \citep{Holm1979} procedure. We used an RBF kernel for the SVM, measured performance via MSE, and used 5 subsampling iterations to evaluate empirical risk. The results are shown in Fig.~\ref{fig:boston}. We found significant effects at $\alpha = 0.05$ for the average number of rooms (\texttt{rm}), percentage of lower status of the population (\texttt{lstat}), pupil-teacher ratio (\texttt{ptratio}), and several other variables with both LM and SVM, which is in line with previous analyses \citep{Friedman2008,Williamson2017}. Interestingly, the SVM assigned a higher CPI value to several variables compared to the LM. For example, the proportion of owner-occupied units built prior to 1940 (\texttt{age}) significantly increased the predictive performance of the SVM but had approximately zero impact on the LM. The reason for this difference might be a nonlinear interaction between \texttt{rm} and \texttt{age}, which was also observed by \citet{Friedman2008}.

\subsubsection{Breast Cancer}\label{sec:pathways}
We examined gene expression profiles of human breast cancer samples downloaded from GEO series GSE3165. Only the 94 arrays of platform GPL887 (Agilent Human 1A Microarray V2) were included. These data were originally analyzed by \citet{Herschkowitz2007} and later studied by \citet{Lim2009}. We follow their preprocessing pipeline, leaving 13,064 genes. All samples were taken from tumor tissue and classified into one of six molecular subtypes: basal-like, luminal A, luminal B, Her2, normal-like, and claudin-low.

Basal-like breast cancer (BLBC) is an especially aggressive form of the disease, and BLBC patients generally have a poor prognosis. Following \citet{Wu2012}, we define a binary response vector to indicate whether or not samples are BLBC. Gene sets were downloaded from the curated C2 collection of the MSigDB and tested for their association with this dichotomous outcome. 

We trained an RF classifier with 10,000 trees to predict BLBC based on microarray data. Second-order knockoffs were sampled using an approximate semidefinite program with block-diagonal covariance matrices of maximum dimension $4,000 \times 4,000$. We test the CPI for each of the 2,609 gene sets in the C2 collection for which at least 25 genes were present in the expression matrix. Models were evaluated using the CE loss function on out-of-bag samples.

\begin{figure}[!h]
\begin{center}
\includegraphics[width=\textwidth]{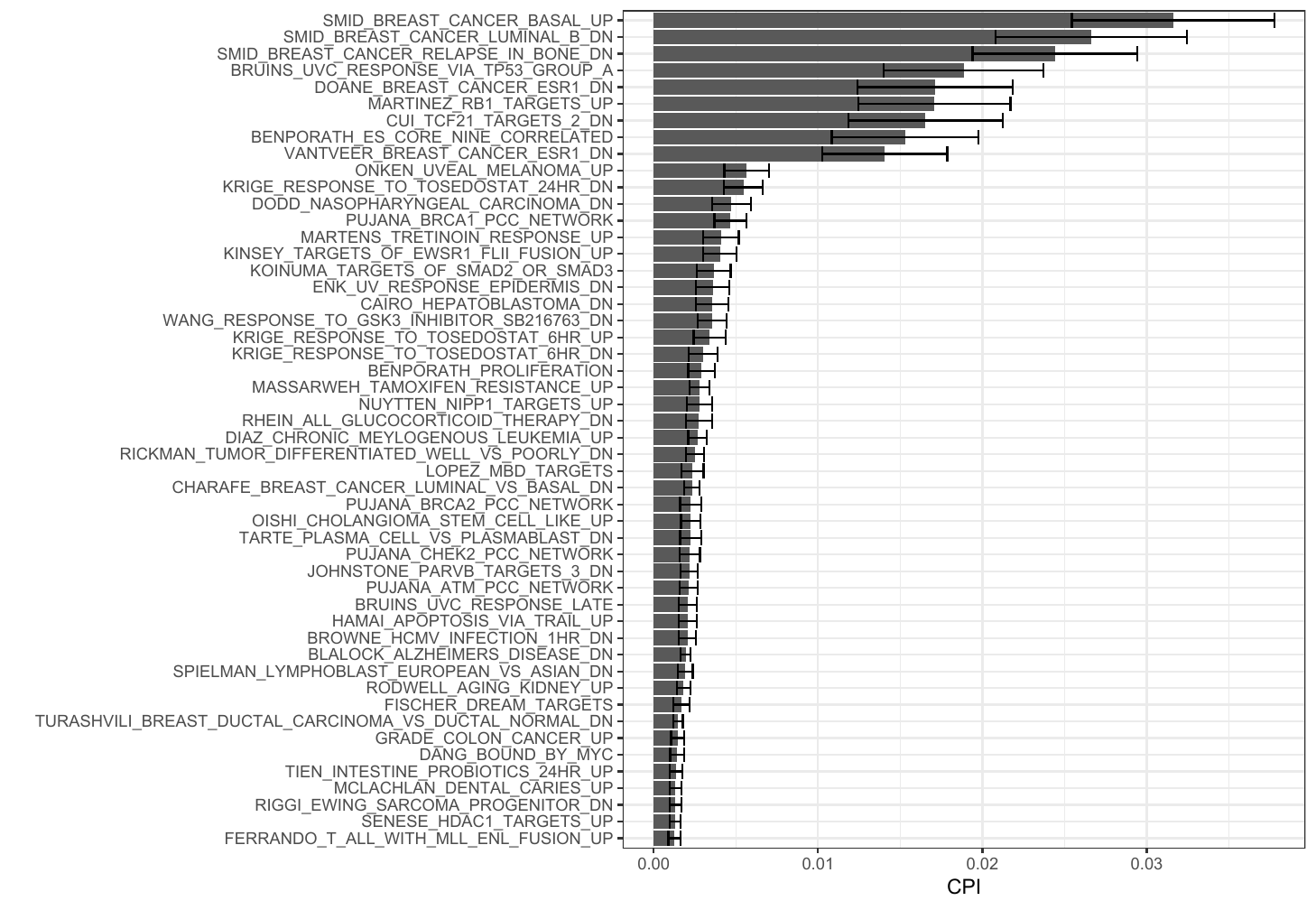}
\end{center}
\caption{Results for the top 50 gene sets. For each gene set, the CPI value is shown, computed with a random forest. Whiskers represent standard errors.}
\label{fig:cancer}
\end{figure}

We calculate $p$-values for each CPI via the $t$-test and corresponding $q$-values using the Benjamini-Hochberg procedure \citep{Benjamini1995}. We identify 660 significantly enriched gene sets at $q \leq 0.05$, including 24 of 73 explicitly breast cancer derived gene sets and 6 of 13 gene sets indicative of basal signatures. Almost all top results are from cancer studies or other biologically relevant research (see Fig.~\ref{fig:cancer}). These include 4 sets of BRCA1 targets, genetic mutations known to be associated with BLBC \citep{Turner2006}, and 4 sets of ESR1 targets, which are known markers for the luminal A subtype \citep{Sorlie2003}. 

By comparison, popular pathway enrichment tests like GSEA \citep{Subramanian2005} and CAMERA \citep{Wu2012} respectively identify 137 and 74 differentially enriched pathways in this dataset at 5\% FDR. Our results are especially notable given that those methods rely on marginal associations between gene expression and clinical outcomes, whereas the CPI is a conditional test with a more restrictive null hypothesis, and should theoretically have less power to detect enrichment when features within a gene set are correlated with others outside it. Despite collinearity between genes, the CPI still identifies a large number of biologically meaningful gene sets differentiating BLBC tumors from other breast cancer subtypes. 

\section{Discussion}\label{sec:discussion}
\citet{Shah2018} have demonstrated that no CI test can be uniformly valid against arbitrary alternatives, a sort of no-free-lunch (NFL) theorem for CI. \citet{Bates2019} prove a similar NFL theorem for constructing knockoff variables, showing that no algorithm can efficiently compute nontrivial knockoffs for arbitrary input distributions. The original NFL theorem for optimization is well-known \citep{Wolpert1997}. Together, these results delimit the scope of the CPI. Our method is completely general, in the sense that it works with any well-chosen combination of supervised learner, loss function, and knockoff sampler. However, it is simultaneously constrained by these choices. The CPI will not in general control Type I error or have any power against the null when knockoffs are poorly constructed or models are misspecified. If the selected loss function ignored distributional information from higher moments, then the CPI will as well. With likelihood-based measures, however, the method will converge on the true Markov blanket.

In our experiments, we employed a variety of risk estimators, including cross-validation, subsampling, out-of-bag estimates, and the holdout method. Results did not depend on these choices, suggesting that analysts may use whichever is most efficient for the problem at hand. We also used a number of different learning algorithms and found that all showed good convergence properties in finite samples -- even the SVM, which is known to have infinite VC dimension, and therefore violates the consistency criterion cited in Sect.~\ref{sec:theory}.

Computational bottlenecks can complicate the use of our procedure for high-dimensional datasets. It took approximately 49 hours to generate second-order knockoffs for the gene expression matrix described in Sect.~\ref{sec:pathways}. However, as noted in Sect.~\ref{sec:ko}, knockoff sampling is an active area of research, and we expect future advances to speed up the procedure considerably.

\section{Conclusion}\label{sec:conclusion}
We propose the conditional predictive impact (CPI), a maximally general test of conditional independence. It works for regression and classification problems using any combination of knockoff sampler, supervised learning algorithm, and loss function. It imposes no parametric or sparsity constraints, and can be efficiently computed on data with many observations and features. Our inference procedures are fast and powerful, able to simultaneously control Type I error and achieve nominal coverage probability. We have shown that our approach is consistent and unbiased under minimal assumptions. Empirical results demonstrate that our method performs favorably against a number of alternatives for a range of supervised learners and data generating processes.

We envision several avenues for future research in this area. Localized versions of the CPI algorithm could be used to detect the conditional importance of features on particular predictions. Model-specific methods could be implemented to speed up the procedure. We are currently working on applications for causal discovery and inference, an especially promising direction for this approach.


\begin{acknowledgements}
We thank our anonymous reviewers for their helpful feedback. DSW received funding for this project from ONR grant N62909-19-1-2096. MNW received funding for this project from the German Research Foundation (DFG), Emmy Noether Grant 437611051. 
\end{acknowledgements}


\bibliographystyle{spbasic}
\bibliography{cpi}

\end{document}


\maketitle

\renewcommand*{\thefootnote}{\fnsymbol{footnote}}

\footnotetext[1]{Corresponding author, email: david.watson@ucl.ac.uk}

\makeatletter 
\renewcommand{\thetable}{S\@arabic\c@table}
\makeatother

\makeatletter 
\renewcommand{\thefigure}{S\@arabic\c@figure}
\makeatother

\begin{table}[htbp]
\centering
\begin{tabular}{lrrrr}
  \toprule
Learner & \multicolumn{2}{c}{Linear data} &  \multicolumn{2}{c}{Non-linear data} \\ 
 & $t$-Test & Fisher & $t$-Test & Fisher \\
  \midrule
  Linear model & 0.9514 & 0.9485 & 0.9533 & 0.9479 \\ 
  Support vector machine & 0.9516 & 0.9550 & 0.9528 & 0.9527 \\ 
  Random forest & 0.9505 & 0.9537 & 0.9546 & 0.9529 \\ 
  Neural network & 0.9518 & 0.9533 & 0.9548 & 0.9501 \\ 
   \bottomrule
\end{tabular}
\caption{Empirical coverage probabilities of $95\%$ confidence intervals in the simulation study, calculated from $10,000$ simulation replicates; continuous outcome with MSE loss function; correlated predictors.} 
\end{table}

\begin{table}[htbp]
\centering
\begin{tabular}{lrrrr}
  \toprule
Learner & \multicolumn{2}{c}{Linear data} &  \multicolumn{2}{c}{Non-linear data} \\ 
 & $t$-Test & Fisher & $t$-Test & Fisher \\
  \midrule
  Linear model & 0.9523 & 0.9496 & 0.9517 & 0.9531 \\ 
  Support vector machine & 0.9517 & 0.9491 & 0.9521 & 0.9539 \\ 
  Random forest & 0.9500 & 0.9516 & 0.9498 & 0.9495 \\ 
  Neural network & 0.9557 & 0.9544 & 0.9521 & 0.9532 \\ 
   \bottomrule
\end{tabular}
\caption{Empirical coverage probabilities of $95\%$ confidence intervals in the simulation study, calculated from $10,000$ simulation replicates; continuous outcome with MAE loss function; correlated predictors.} 
\end{table}

\begin{table}[htbp]
\centering
\begin{tabular}{lrrrr}
  \toprule
Learner & \multicolumn{2}{c}{Linear data} &  \multicolumn{2}{c}{Non-linear data} \\ 
 & $t$-Test & Fisher & $t$-Test & Fisher \\
  \midrule
  Linear model & 0.9494 & 0.9484 & 0.9549 & 0.9514 \\ 
  Support vector machine & 0.9514 & 0.9544 & 0.9518 & 0.9542 \\ 
  Random forest & 0.9533 & 0.9544 & 0.9503 & 0.9537 \\ 
  Neural network & 0.9537 & 0.9525 & 0.9493 & 0.9538 \\ 
   \bottomrule
\end{tabular}
\caption{Empirical coverage probabilities of $95\%$ confidence intervals in the simulation study, calculated from $10,000$ simulation replicates; continuous outcome with MSE loss function; uncorrelated predictors.} 
\end{table}

\begin{table}[htbp]
\centering
\begin{tabular}{lrrrr}
  \toprule
Learner & \multicolumn{2}{c}{Linear data} &  \multicolumn{2}{c}{Non-linear data} \\ 
 & $t$-Test & Fisher & $t$-Test & Fisher \\
  \midrule
  Linear model & 0.9522 & 0.9486 & 0.9525 & 0.9502 \\ 
  Support vector machine & 0.9511 & 0.9515 & 0.9513 & 0.9534 \\ 
  Random forest & 0.9518 & 0.9511 & 0.9509 & 0.9517 \\ 
  Neural network & 0.9506 & 0.9470 & 0.9508 & 0.9538 \\ 
   \bottomrule
\end{tabular}
\caption{Empirical coverage probabilities of $95\%$ confidence intervals in the simulation study, calculated from $10,000$ simulation replicates; continuous outcome with MAE loss function; uncorrelated predictors.} 
\end{table}

\begin{table}[htbp]
\centering
\begin{tabular}{lrrrr}
  \toprule
Learner & \multicolumn{2}{c}{Linear data} &  \multicolumn{2}{c}{Non-linear data} \\ 
 & $t$-Test & Fisher & $t$-Test & Fisher \\
  \midrule
  Logistic regression & 0.7405 & 0.9576 & 0.9029 & 0.9493 \\ 
  Support vector machine & 0.9034 & 0.9525 & 0.9260 & 0.9503 \\ 
  Random forest & 0.8585 & 0.9473 & 0.9080 & 0.9454 \\ 
  Neural network & 0.9303 & 0.9529 & 0.9357 & 0.9519 \\ 
   \bottomrule
\end{tabular}
\caption{Empirical coverage probabilities of $95\%$ confidence intervals in the simulation study, calculated from $10,000$ simulation replicates; classification outcome with MMCE loss function; correlated predictors.} 
\end{table}

\begin{table}[htbp]
\centering
\begin{tabular}{lrrrr}
  \toprule
Learner & \multicolumn{2}{c}{Linear data} &  \multicolumn{2}{c}{Non-linear data} \\ 
 & $t$-Test & Fisher & $t$-Test & Fisher \\
  \midrule
  Logistic regression & 0.9511 & 0.9516 & 0.9510 & 0.9511 \\ 
  Support vector machine & 0.9514 & 0.9520 & 0.9534 & 0.9495 \\ 
  Random forest & 0.9503 & 0.9499 & 0.9521 & 0.9503 \\ 
  Neural network & 0.9493 & 0.9522 & 0.9517 & 0.9518 \\ 
   \bottomrule
\end{tabular}
\caption{Empirical coverage probabilities of $95\%$ confidence intervals in the simulation study, calculated from $10,000$ simulation replicates; classification outcome with CE loss function; correlated predictors.} 
\end{table}

\begin{table}[htbp]
\centering
\begin{tabular}{lrrrr}
  \toprule
Learner & \multicolumn{2}{c}{Linear data} &  \multicolumn{2}{c}{Non-linear data} \\ 
 & $t$-Test & Fisher & $t$-Test & Fisher \\
  \midrule
  Logistic regression & 0.7651 & 0.9525 & 0.8968 & 0.9523 \\ 
  Support vector machine & 0.9164 & 0.9503 & 0.9262 & 0.9532 \\ 
  Random forest & 0.9023 & 0.9439 & 0.9110 & 0.9507 \\ 
  Neural network & 0.9345 & 0.9504 & 0.9369 & 0.9514 \\ 
   \bottomrule
\end{tabular}
\caption{Empirical coverage probabilities of $95\%$ confidence intervals in the simulation study, calculated from $10,000$ simulation replicates; classification outcome with MMCE loss function; uncorrelated predictors.} 
\end{table}

\begin{table}[htbp]
\centering
\begin{tabular}{lrrrr}
  \toprule
Learner & \multicolumn{2}{c}{Linear data} &  \multicolumn{2}{c}{Non-linear data} \\ 
 & $t$-Test & Fisher & $t$-Test & Fisher \\
  \midrule
Logistic regression & 0.9510 & 0.9480 & 0.9487 & 0.9469 \\ 
  Support vector machine & 0.9488 & 0.9492 & 0.9527 & 0.9493 \\ 
  Random forest & 0.9532 & 0.9524 & 0.9499 & 0.9516 \\ 
  Neural network & 0.9522 & 0.9523 & 0.9494 & 0.9522 \\ 
   \bottomrule
\end{tabular}
\caption{Empirical coverage probabilities of $95\%$ confidence intervals in the simulation study, calculated from $10,000$ simulation replicates; classification outcome with CE loss function; uncorrelated predictors.} 
\end{table}

\begin{figure}[!ht]
\centering
\includegraphics[width=\textwidth]{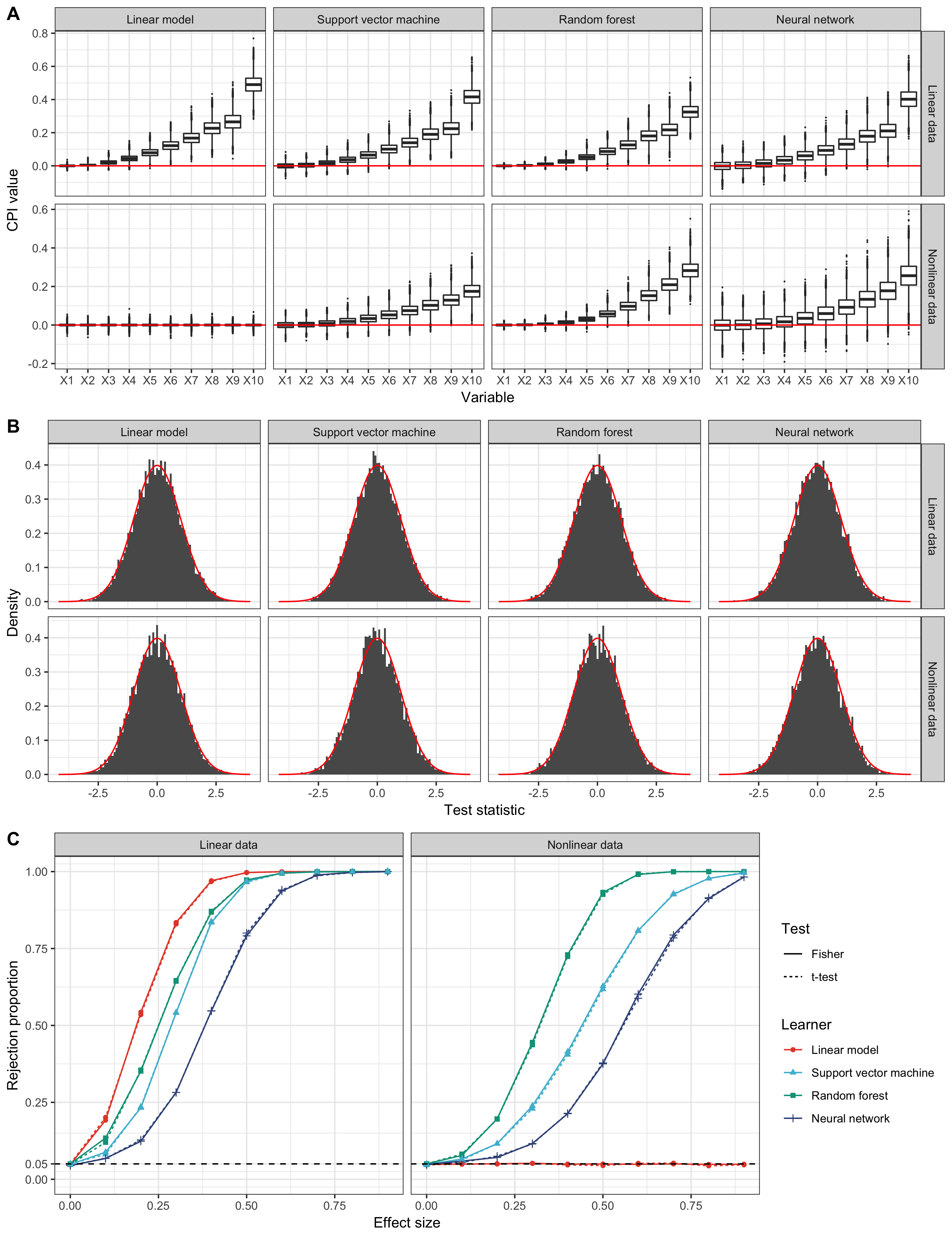}
\caption{Simulation results for continuous outcome with MAE loss function and correlated predictors. \textbf{A}: Boxplots of simulation replications of CPI values of variables $X_1, \dots, X_{10}$ with increasing effect size. The red line indicates a CPI value of 0, corresponding to no estimated association between the variable $X_j$ and the outcome $Y$. \textbf{B}: Histograms of simulation replications of $t$-statistics of variables with effect size $0$. The distribution of the expected $t$-statistic under the null hypothesis is shown in red. \textbf{C}: Average proportion of rejected hypotheses at $\alpha = 0.05$. Results at effect size $0$ correspond to the type I error, at effect sizes $>0$ to statistical power. The dashed line indicates the nominal level of $\alpha = 0.05$. The panels correspond to the simulation scenario, the colors and symbols to the learning algorithms and the line types to the inference procedure.}
\end{figure}

\begin{figure}[!ht]
\centering
\includegraphics[width=\textwidth]{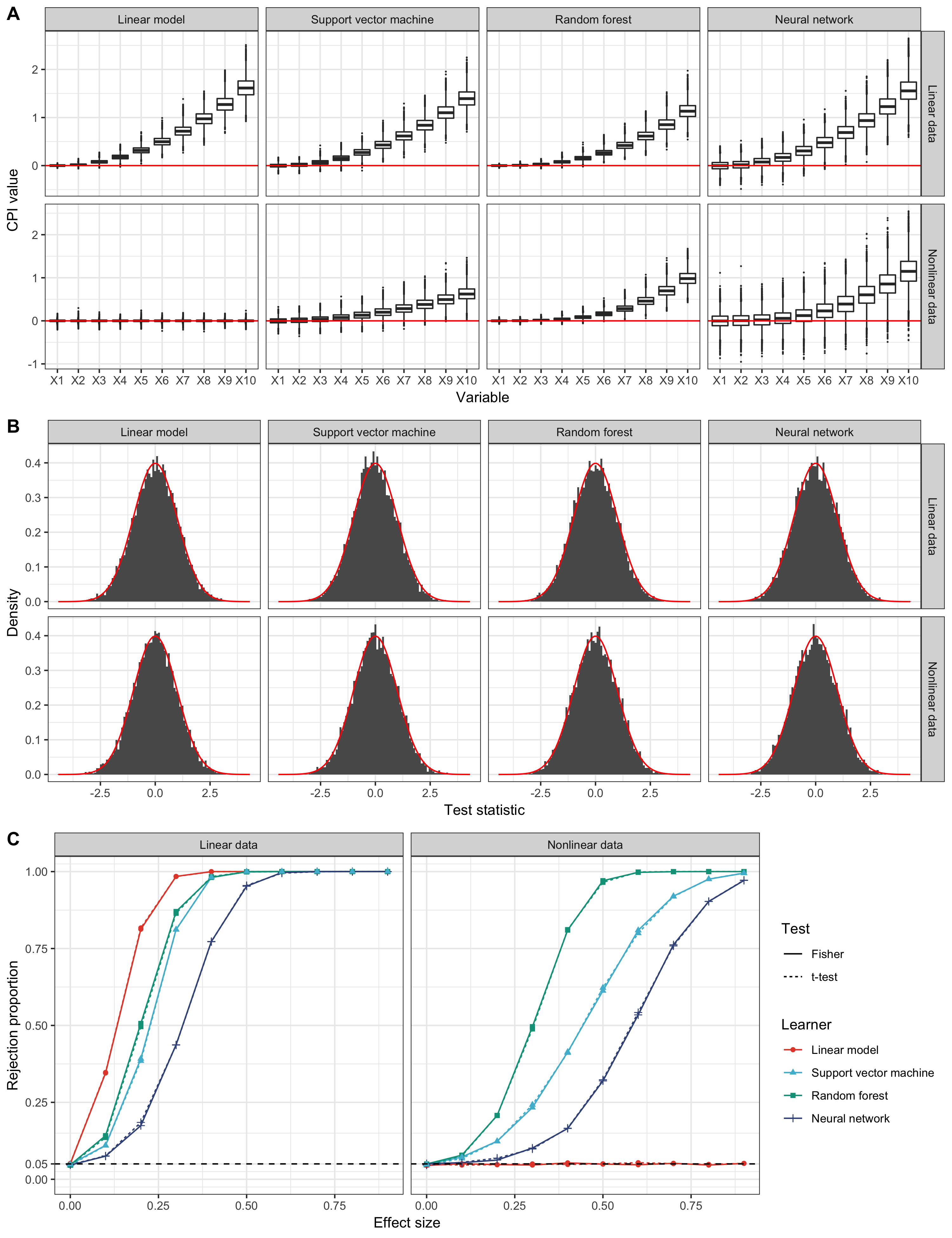}
\caption{Simulation results for continuous outcome with MSE loss function and uncorrelated predictors. \textbf{A}: Boxplots of simulation replications of CPI values of variables $X_1, \dots, X_{10}$ with increasing effect size. The red line indicates a CPI value of 0, corresponding to no estimated association between the variable $X_j$ and the outcome $Y$. \textbf{B}: Histograms of simulation replications of $t$-statistics of variables with effect size $0$. The distribution of the expected $t$-statistic under the null hypothesis is shown in red. \textbf{C}: Average proportion of rejected hypotheses at $\alpha = 0.05$. Results at effect size $0$ correspond to the type I error, at effect sizes $>0$ to statistical power. The dashed line indicates the nominal level of $\alpha = 0.05$. The panels correspond to the simulation scenario, the colors and symbols to the learning algorithms and the line types to the inference procedure.}
\end{figure}

\begin{figure}[!ht]
\centering
\includegraphics[width=\textwidth]{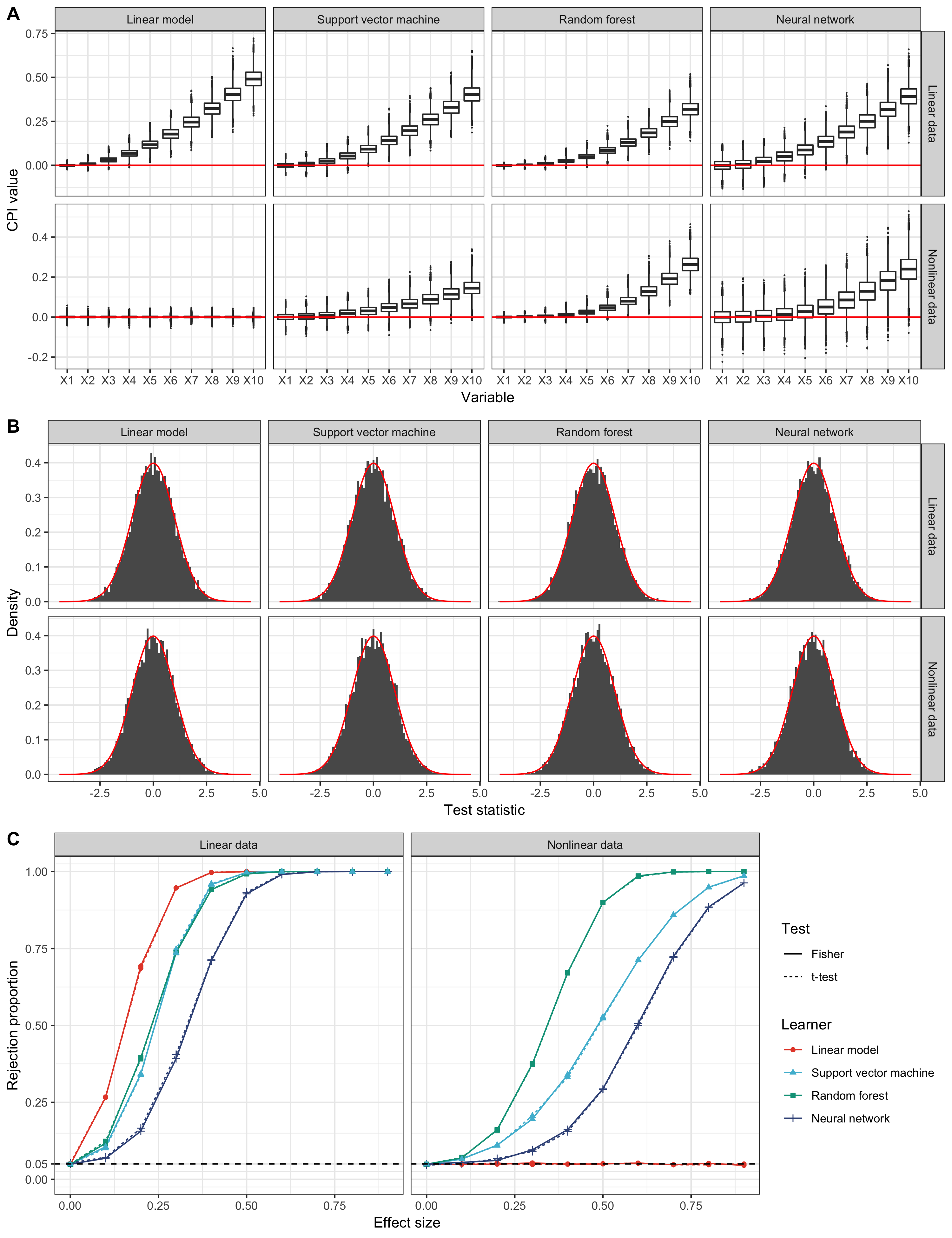}
\caption{Simulation results for continuous outcome with MAE loss function and uncorrelated predictors. \textbf{A}: Boxplots of simulation replications of CPI values of variables $X_1, \dots, X_{10}$ with increasing effect size. The red line indicates a CPI value of 0, corresponding to no estimated association between the variable $X_j$ and the outcome $Y$. \textbf{B}: Histograms of simulation replications of $t$-statistics of variables with effect size $0$. The distribution of the expected $t$-statistic under the null hypothesis is shown in red. \textbf{C}: Average proportion of rejected hypotheses at $\alpha = 0.05$. Results at effect size $0$ correspond to the type I error, at effect sizes $>0$ to statistical power. The dashed line indicates the nominal level of $\alpha = 0.05$. The panels correspond to the simulation scenario, the colors and symbols to the learning algorithms and the line types to the inference procedure.}
\end{figure}

\begin{figure}[!ht]
\centering
\includegraphics[width=\textwidth]{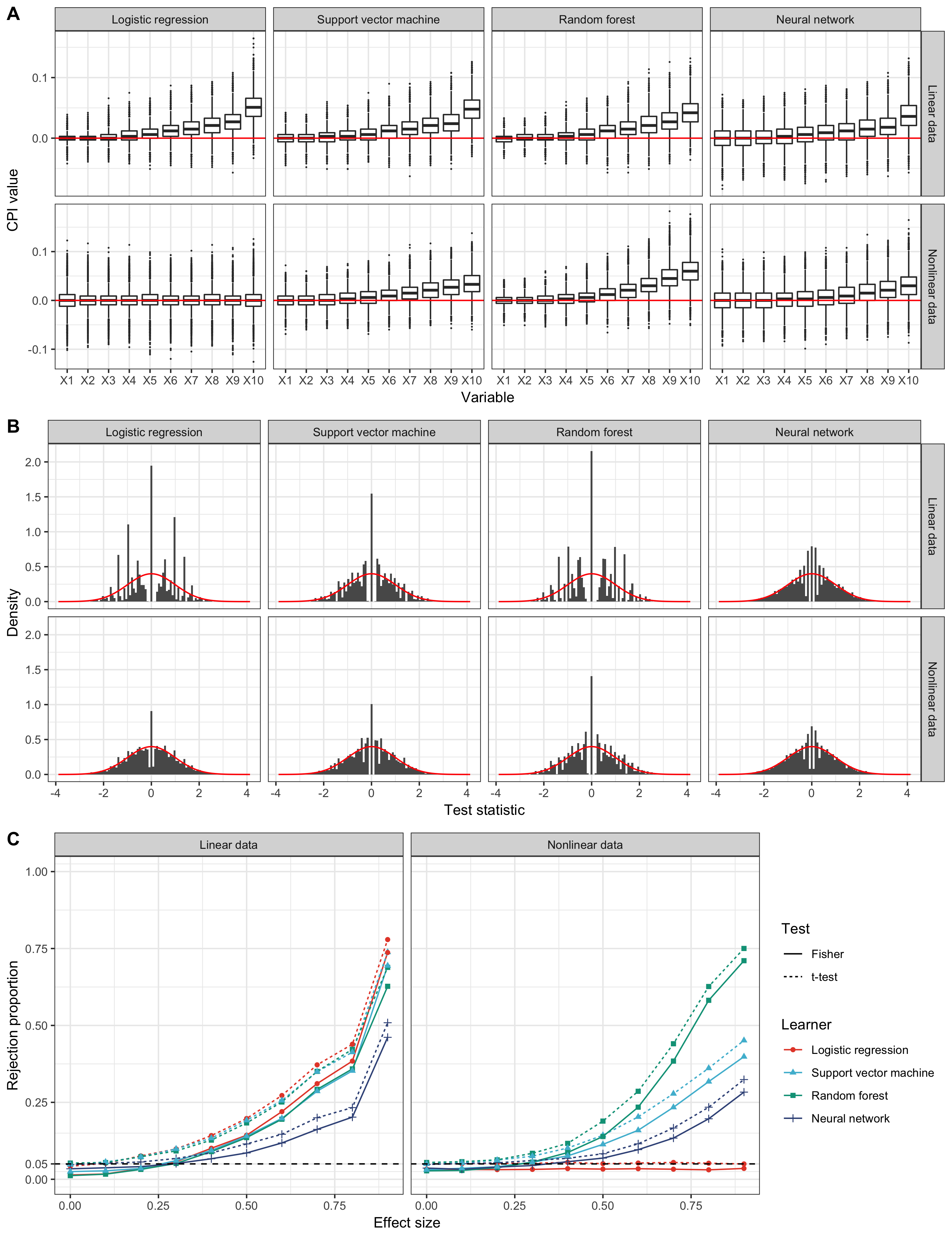}
\caption{Simulation results for classification outcome with MMCE loss function and correlated predictors. \textbf{A}: Boxplots of simulation replications of CPI values of variables $X_1, \dots, X_{10}$ with increasing effect size. The red line indicates a CPI value of 0, corresponding to no estimated association between the variable $X_j$ and the outcome $Y$. \textbf{B}: Histograms of simulation replications of $t$-statistics of variables with effect size $0$. The distribution of the expected $t$-statistic under the null hypothesis is shown in red. \textbf{C}: Average proportion of rejected hypotheses at $\alpha = 0.05$. Results at effect size $0$ correspond to the type I error, at effect sizes $>0$ to statistical power. The dashed line indicates the nominal level of $\alpha = 0.05$. The panels correspond to the simulation scenario, the colors and symbols to the learning algorithms and the line types to the inference procedure.}
\end{figure}

\begin{figure}[!ht]
\centering
\includegraphics[width=\textwidth]{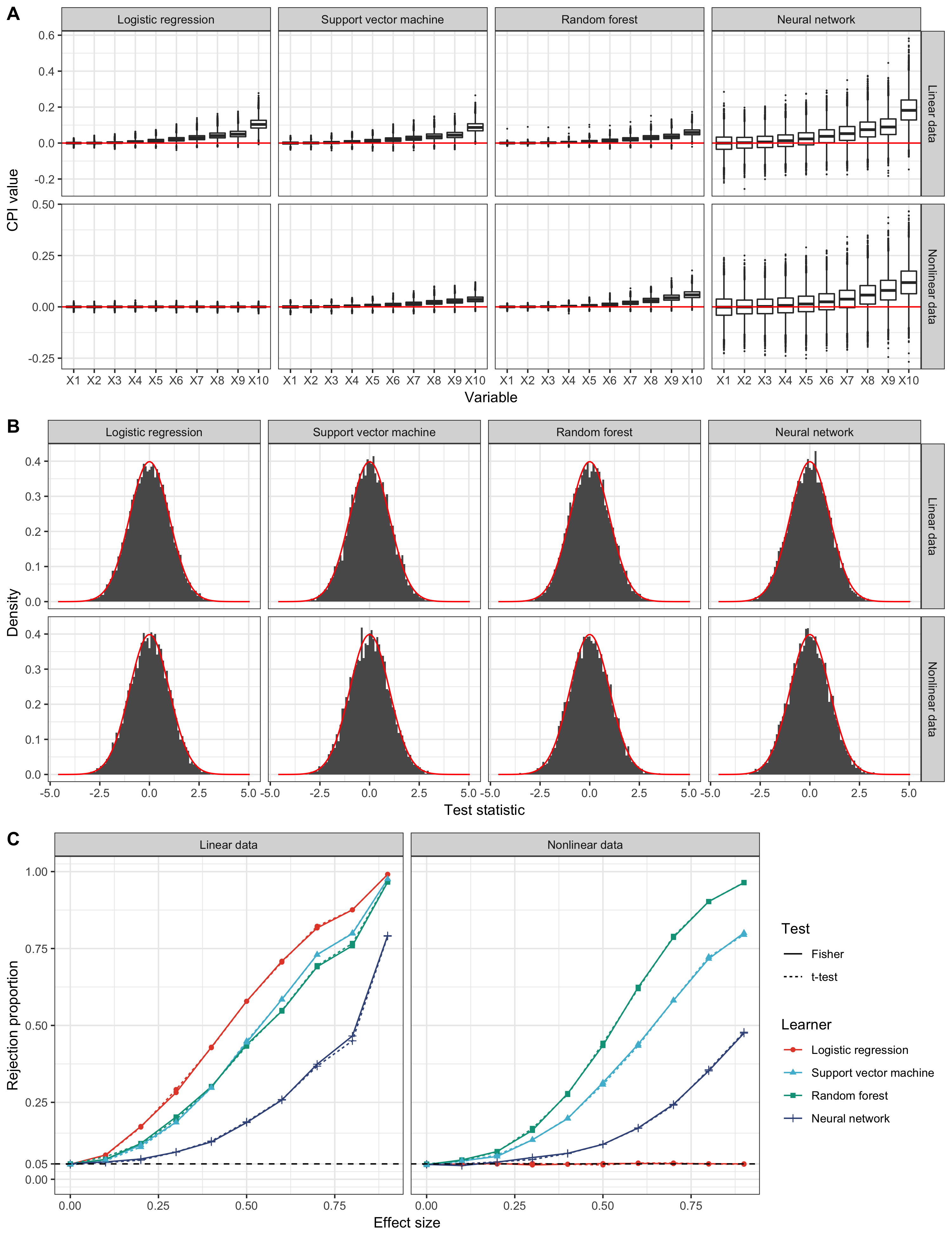}
\caption{Simulation results for classification outcome with CE loss function and correlated predictors. \textbf{A}: Boxplots of simulation replications of CPI values of variables $X_1, \dots, X_{10}$ with increasing effect size. The red line indicates a CPI value of 0, corresponding to no estimated association between the variable $X_j$ and the outcome $Y$. \textbf{B}: Histograms of simulation replications of $t$-statistics of variables with effect size $0$. The distribution of the expected $t$-statistic under the null hypothesis is shown in red. \textbf{C}: Average proportion of rejected hypotheses at $\alpha = 0.05$. Results at effect size $0$ correspond to the type I error, at effect sizes $>0$ to statistical power. The dashed line indicates the nominal level of $\alpha = 0.05$. The panels correspond to the simulation scenario, the colors and symbols to the learning algorithms and the line types to the inference procedure.}
\end{figure}

\begin{figure}[!ht]
\centering
\includegraphics[width=\textwidth]{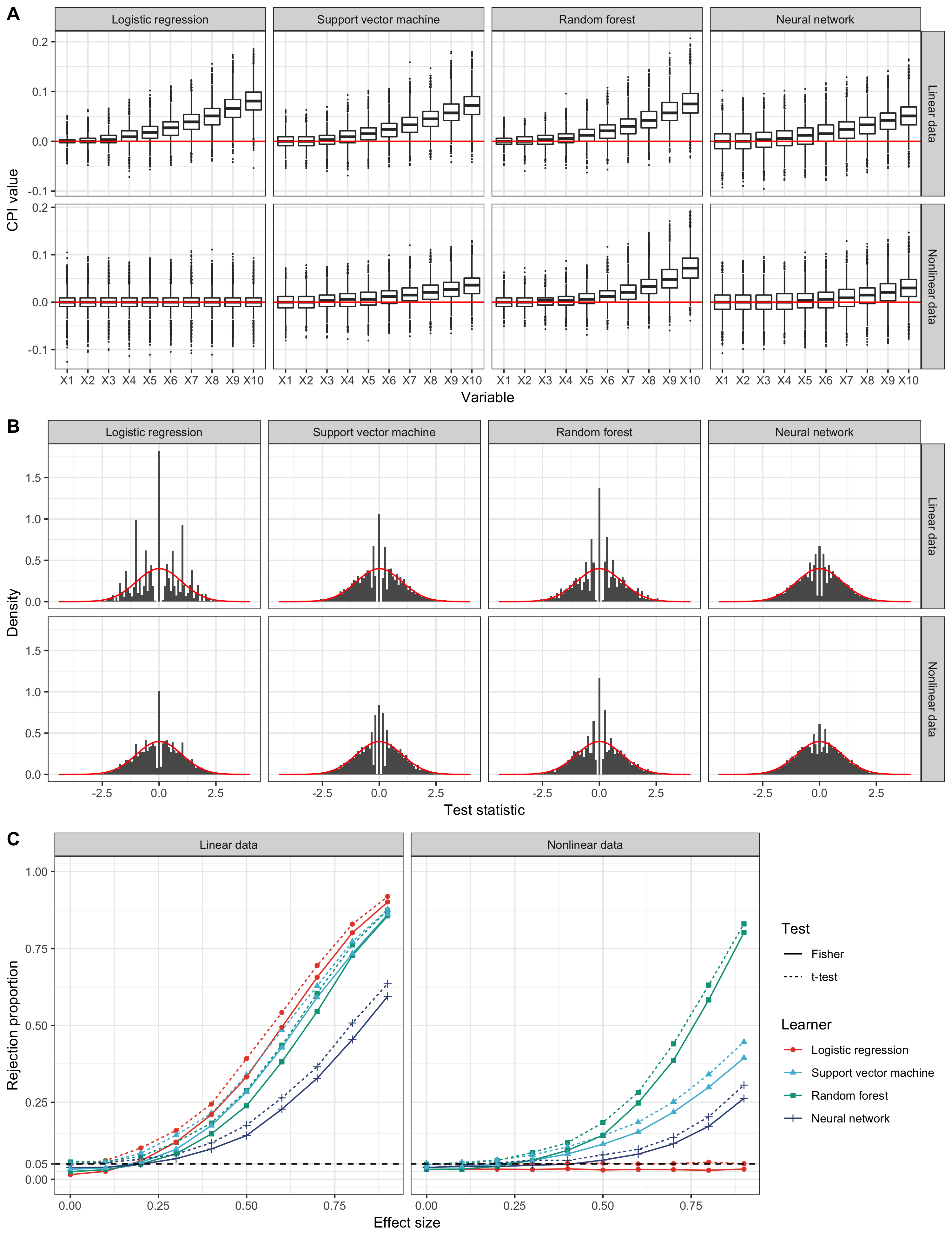}
\caption{Simulation results for classification outcome with MMCE loss function and uncorrelated predictors. \textbf{A}: Boxplots of simulation replications of CPI values of variables $X_1, \dots, X_{10}$ with increasing effect size. The red line indicates a CPI value of 0, corresponding to no estimated association between the variable $X_j$ and the outcome $Y$. \textbf{B}: Histograms of simulation replications of $t$-statistics of variables with effect size $0$. The distribution of the expected $t$-statistic under the null hypothesis is shown in red. \textbf{C}: Average proportion of rejected hypotheses at $\alpha = 0.05$. Results at effect size $0$ correspond to the type I error, at effect sizes $>0$ to statistical power. The dashed line indicates the nominal level of $\alpha = 0.05$. The panels correspond to the simulation scenario, the colors and symbols to the learning algorithms and the line types to the inference procedure.}
\end{figure}

\begin{figure}[!ht]
\centering
\includegraphics[width=\textwidth]{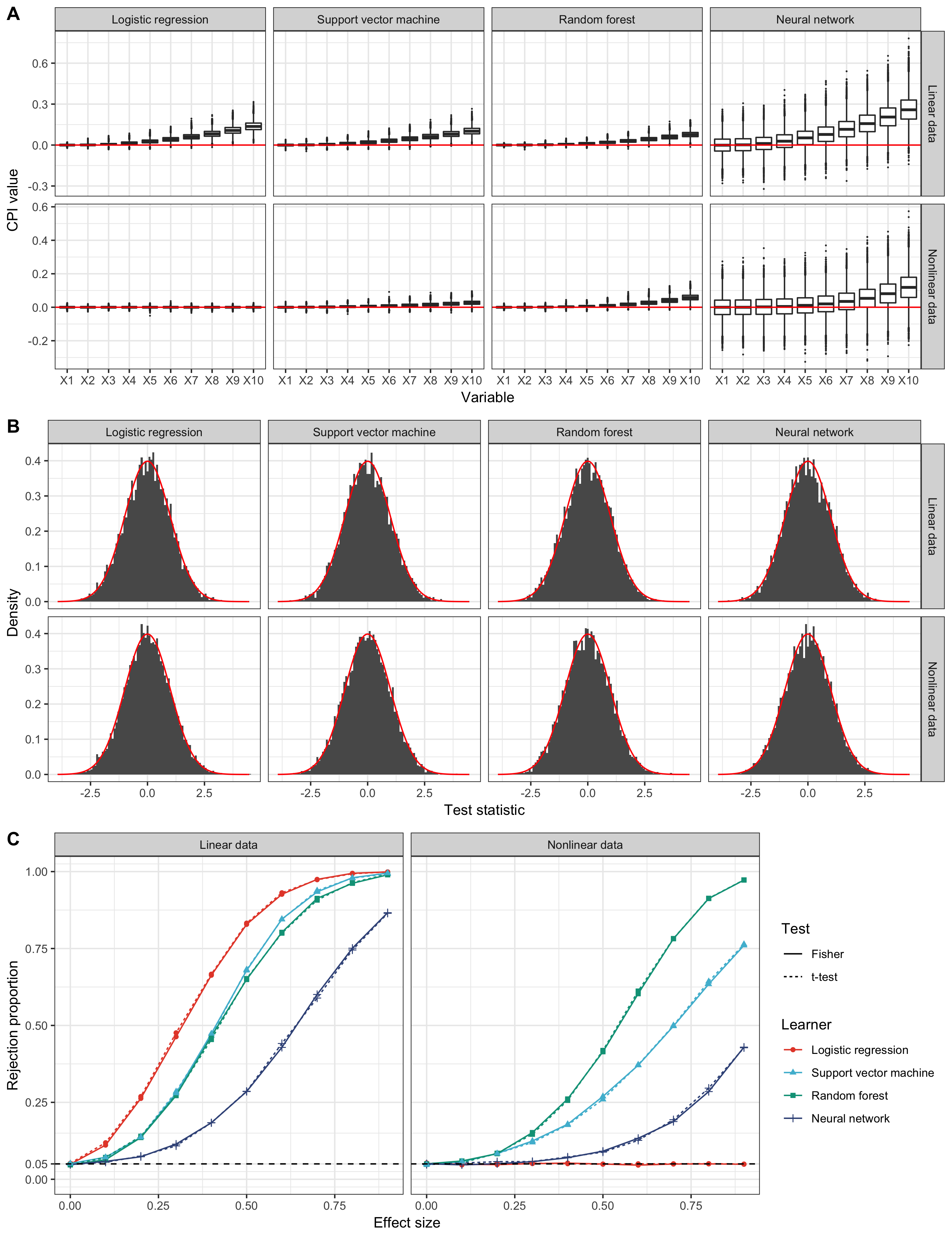}
\caption{Simulation results for classification outcome with CE loss function and uncorrelated predictors. \textbf{A}: Boxplots of simulation replications of CPI values of variables $X_1, \dots, X_{10}$ with increasing effect size. The red line indicates a CPI value of 0, corresponding to no estimated association between the variable $X_j$ and the outcome $Y$. \textbf{B}: Histograms of simulation replications of $t$-statistics of variables with effect size $0$. The distribution of the expected $t$-statistic under the null hypothesis is shown in red. \textbf{C}: Average proportion of rejected hypotheses at $\alpha = 0.05$. Results at effect size $0$ correspond to the type I error, at effect sizes $>0$ to statistical power. The dashed line indicates the nominal level of $\alpha = 0.05$. The panels correspond to the simulation scenario, the colors and symbols to the learning algorithms and the line types to the inference procedure.}
\end{figure}

\begin{figure}[!ht]
\centering
\includegraphics[width=\textwidth]{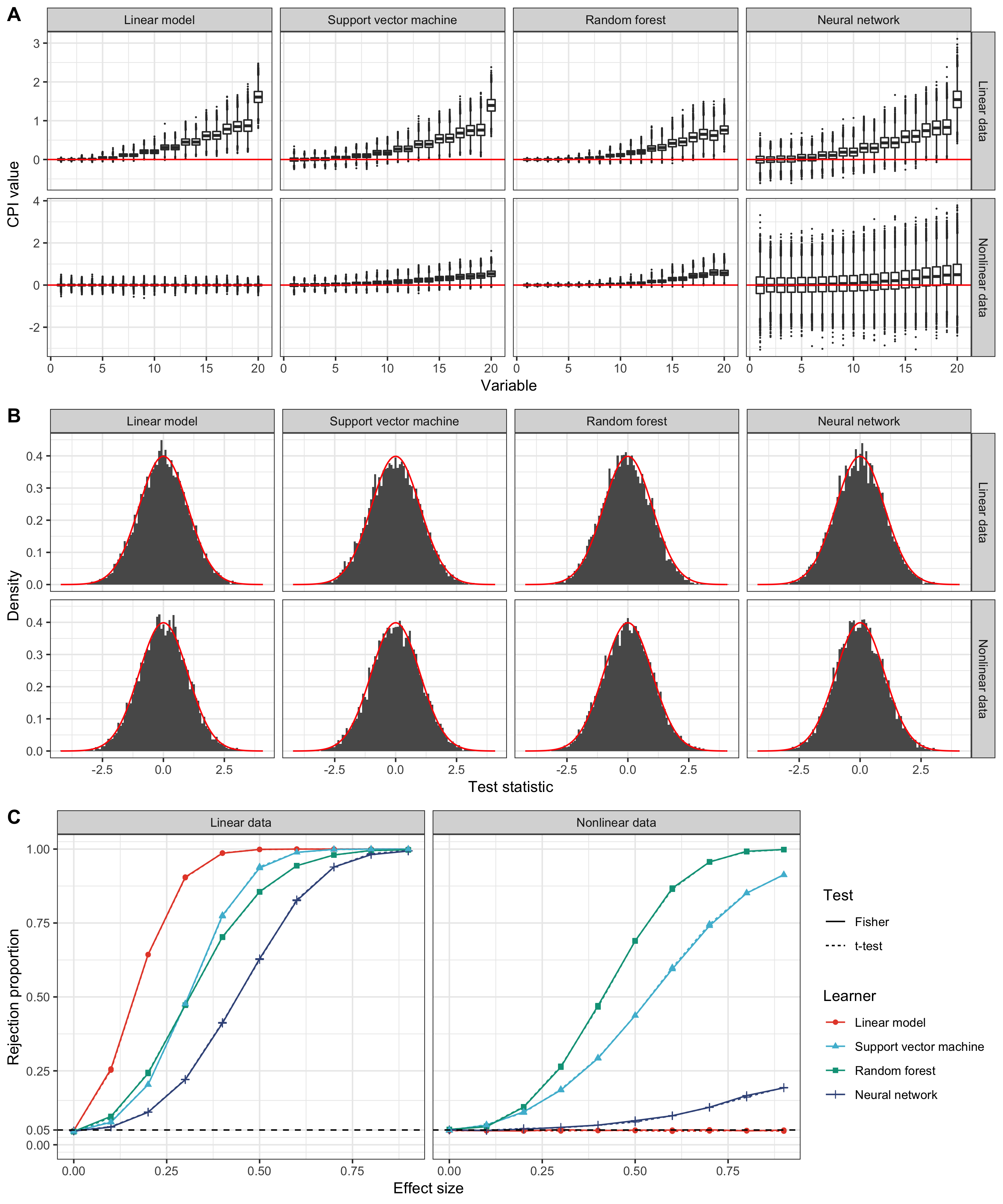}
\caption{Simulation results for continuous outcome with MSE loss function and correlated predictors, $p=20$. \textbf{A}: Boxplots of simulation replications of CPI values of variables $X_1, \dots, X_{20}$ with increasing effect size. The red line indicates a CPI value of 0, corresponding to no estimated association between the variable $X_j$ and the outcome $Y$. \textbf{B}: Histograms of simulation replications of $t$-statistics of variables with effect size $0$. The distribution of the expected $t$-statistic under the null hypothesis is shown in red. \textbf{C}: Average proportion of rejected hypotheses at $\alpha = 0.05$. Results at effect size $0$ correspond to the type I error, at effect sizes $>0$ to statistical power. The dashed line indicates the nominal level of $\alpha = 0.05$. The panels correspond to the simulation scenario, the colors and symbols to the learning algorithms and the line types to the inference procedure.}
\end{figure}

\begin{figure}[!ht]
\centering
\includegraphics[width=\textwidth]{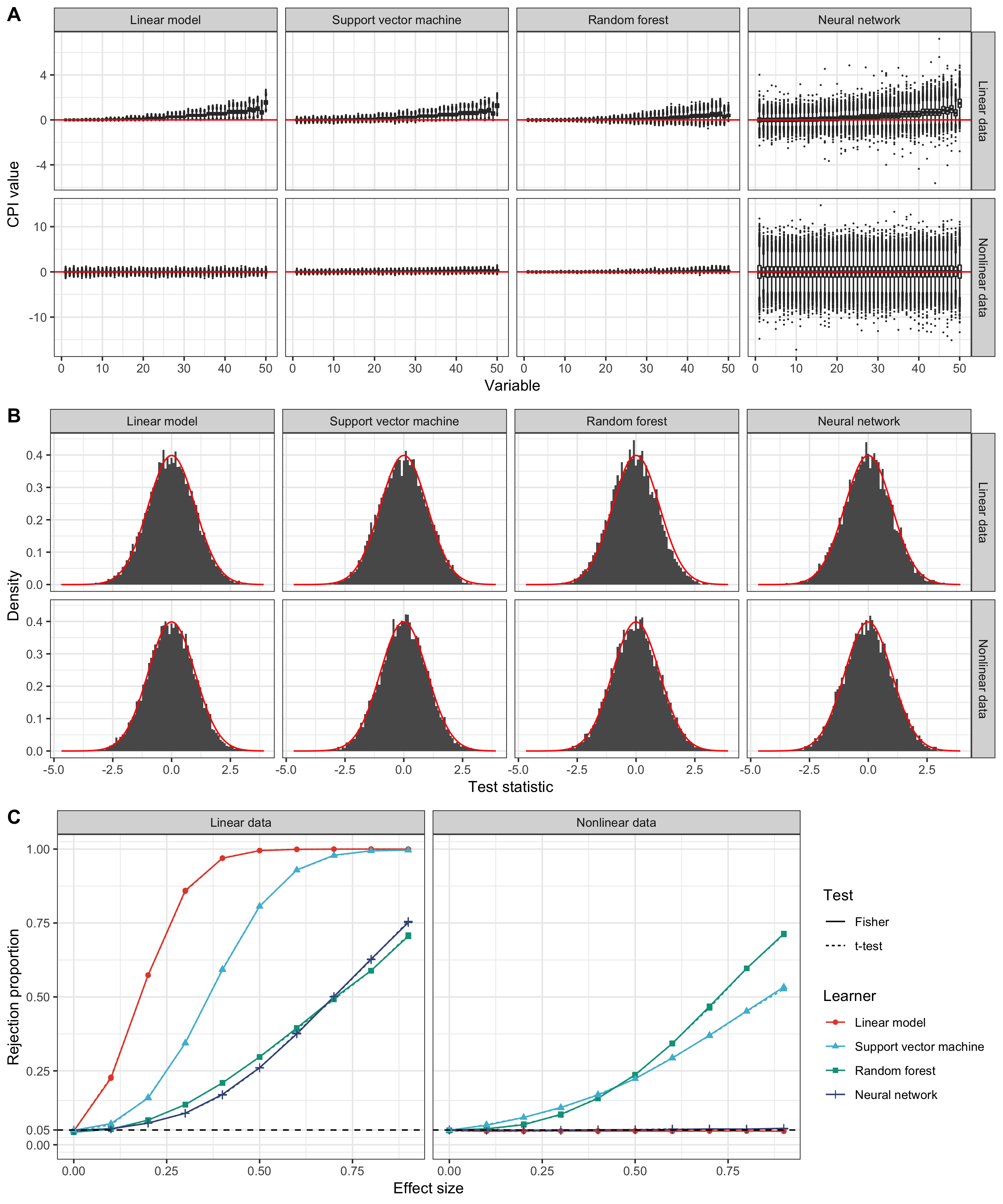}
\caption{Simulation results for continuous outcome with MSE loss function and correlated predictors, $p=50$. \textbf{A}: Boxplots of simulation replications of CPI values of variables $X_1, \dots, X_{50}$ with increasing effect size. The red line indicates a CPI value of 0, corresponding to no estimated association between the variable $X_j$ and the outcome $Y$. \textbf{B}: Histograms of simulation replications of $t$-statistics of variables with effect size $0$. The distribution of the expected $t$-statistic under the null hypothesis is shown in red. \textbf{C}: Average proportion of rejected hypotheses at $\alpha = 0.05$. Results at effect size $0$ correspond to the type I error, at effect sizes $>0$ to statistical power. The dashed line indicates the nominal level of $\alpha = 0.05$. The panels correspond to the simulation scenario, the colors and symbols to the learning algorithms and the line types to the inference procedure.}
\end{figure}

\begin{figure}[!ht]
\centering
\includegraphics[width=\textwidth]{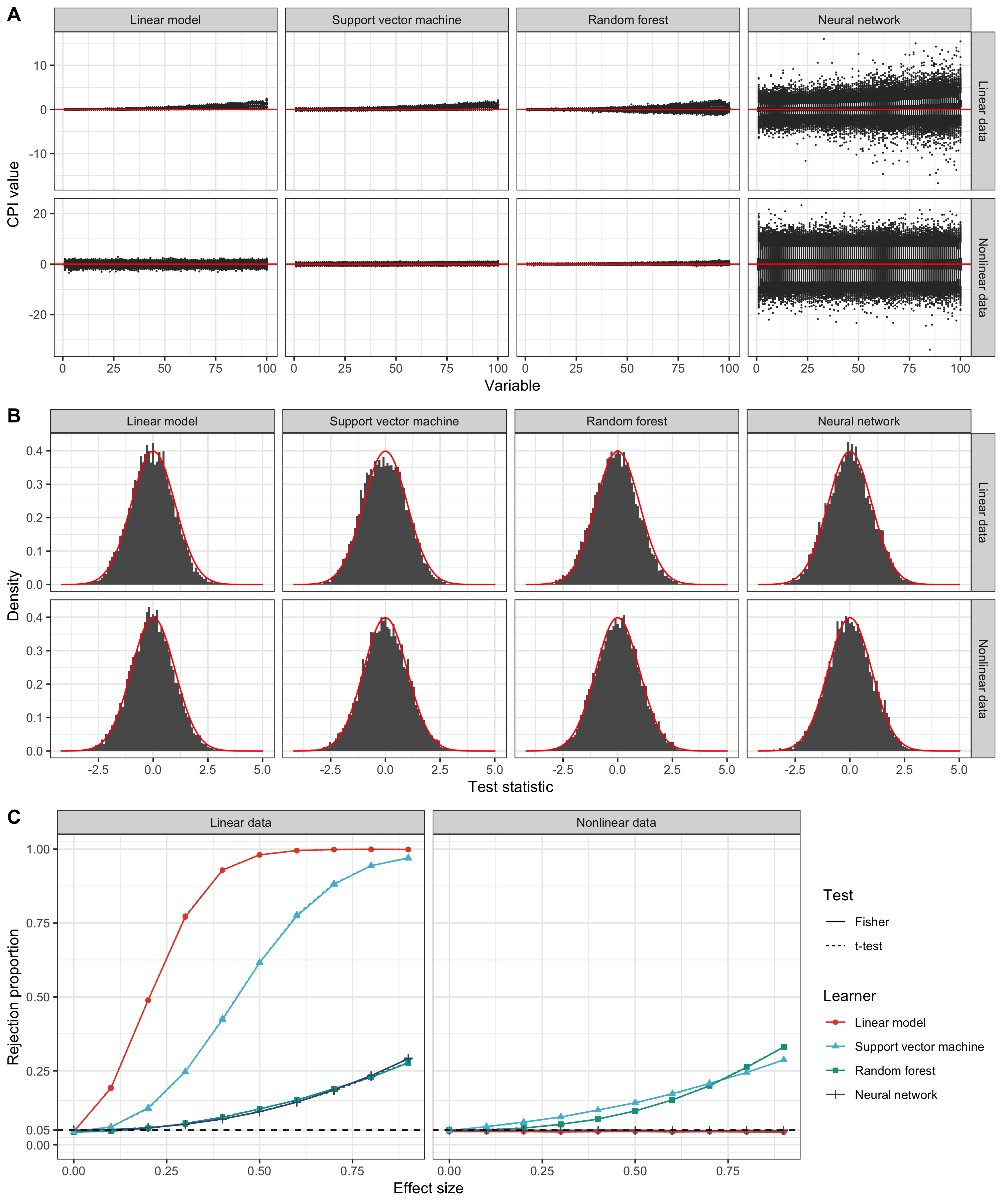}
\caption{Simulation results for continuous outcome with MSE loss function and correlated predictors, $p=100$. \textbf{A}: Boxplots of simulation replications of CPI values of variables $X_1, \dots, X_{100}$ with increasing effect size. The red line indicates a CPI value of 0, corresponding to no estimated association between the variable $X_j$ and the outcome $Y$. \textbf{B}: Histograms of simulation replications of $t$-statistics of variables with effect size $0$. The distribution of the expected $t$-statistic under the null hypothesis is shown in red. \textbf{C}: Average proportion of rejected hypotheses at $\alpha = 0.05$. Results at effect size $0$ correspond to the type I error, at effect sizes $>0$ to statistical power. The dashed line indicates the nominal level of $\alpha = 0.05$. The panels correspond to the simulation scenario, the colors and symbols to the learning algorithms and the line types to the inference procedure.}
\end{figure}

\begin{figure}[htbp]
\begin{centering}
\includegraphics[width=.99\textwidth]{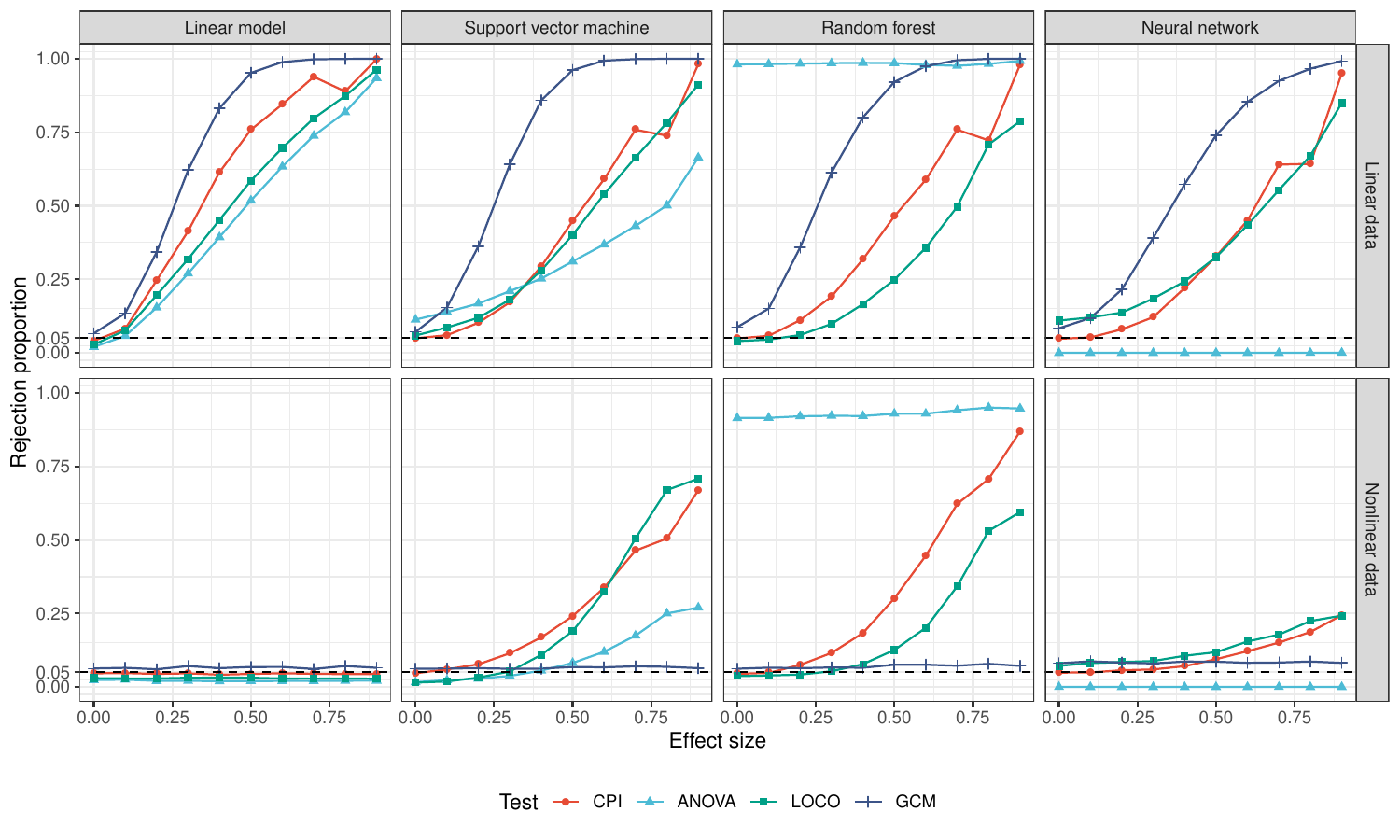}
\end{centering}
\caption{Comparative performance of VI measures across different simulations and algorithms, computed with a training and test sample of $n = 100$, $p = 10$ and correlated predictors. Plots depict the proportion of rejected hypotheses at $\alpha = 0.05$ as a function of effect size. Results at effect size $0$ correspond to Type I error, at effect sizes $>0$ to statistical power. The dashed line indicates the nominal level of $\alpha = 0.05$.}
\end{figure}

\begin{figure}[htbp]
\begin{centering}
\includegraphics[width=.99\textwidth]{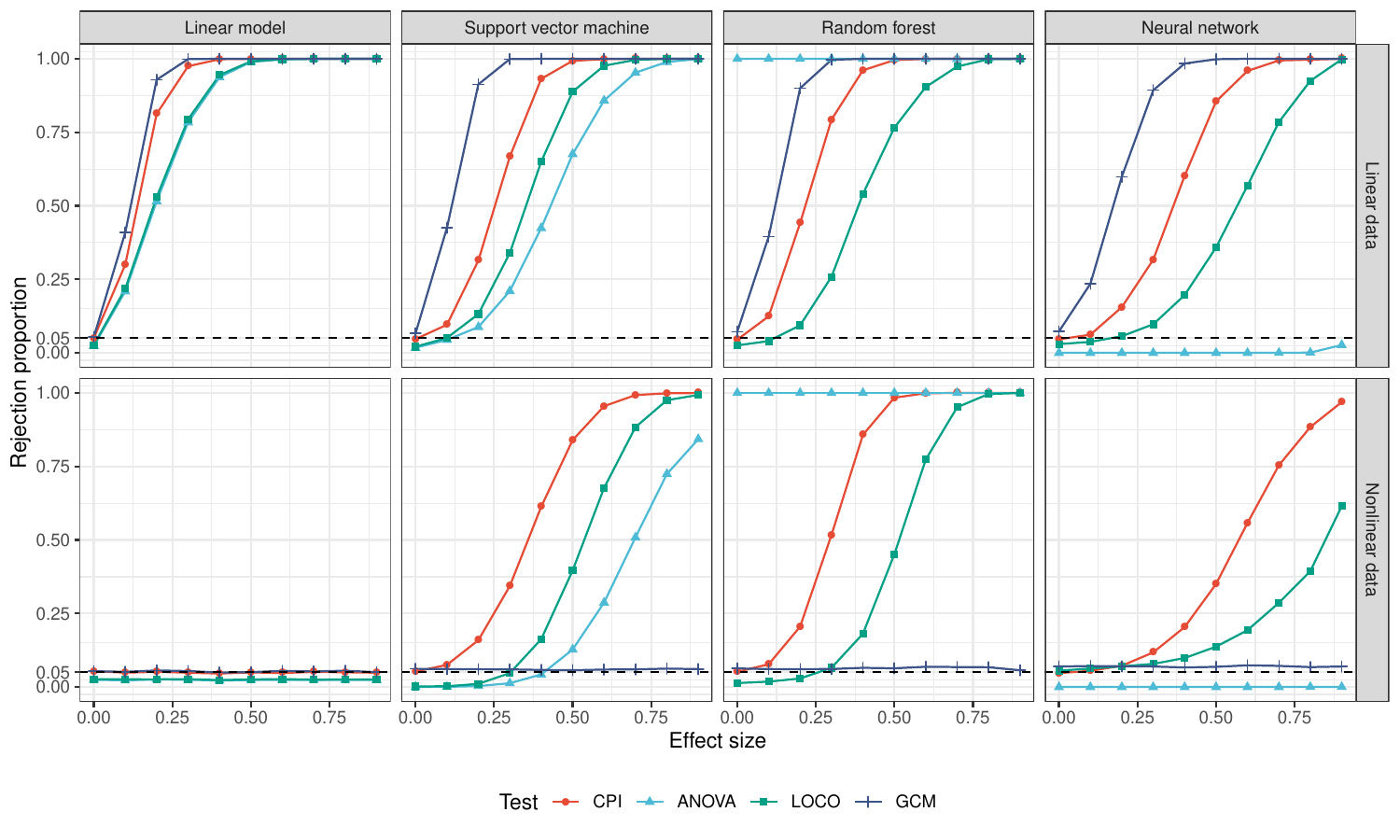}
\end{centering}
\caption{Comparative performance of VI measures across different simulations and algorithms, computed with a training and test sample of $n = 500$, $p = 10$ and correlated predictors. Plots depict the proportion of rejected hypotheses at $\alpha = 0.05$ as a function of effect size. Results at effect size $0$ correspond to Type I error, at effect sizes $>0$ to statistical power. The dashed line indicates the nominal level of $\alpha = 0.05$.}
\end{figure}

\begin{figure}[htbp]
\begin{centering}
\includegraphics[width=.99\textwidth]{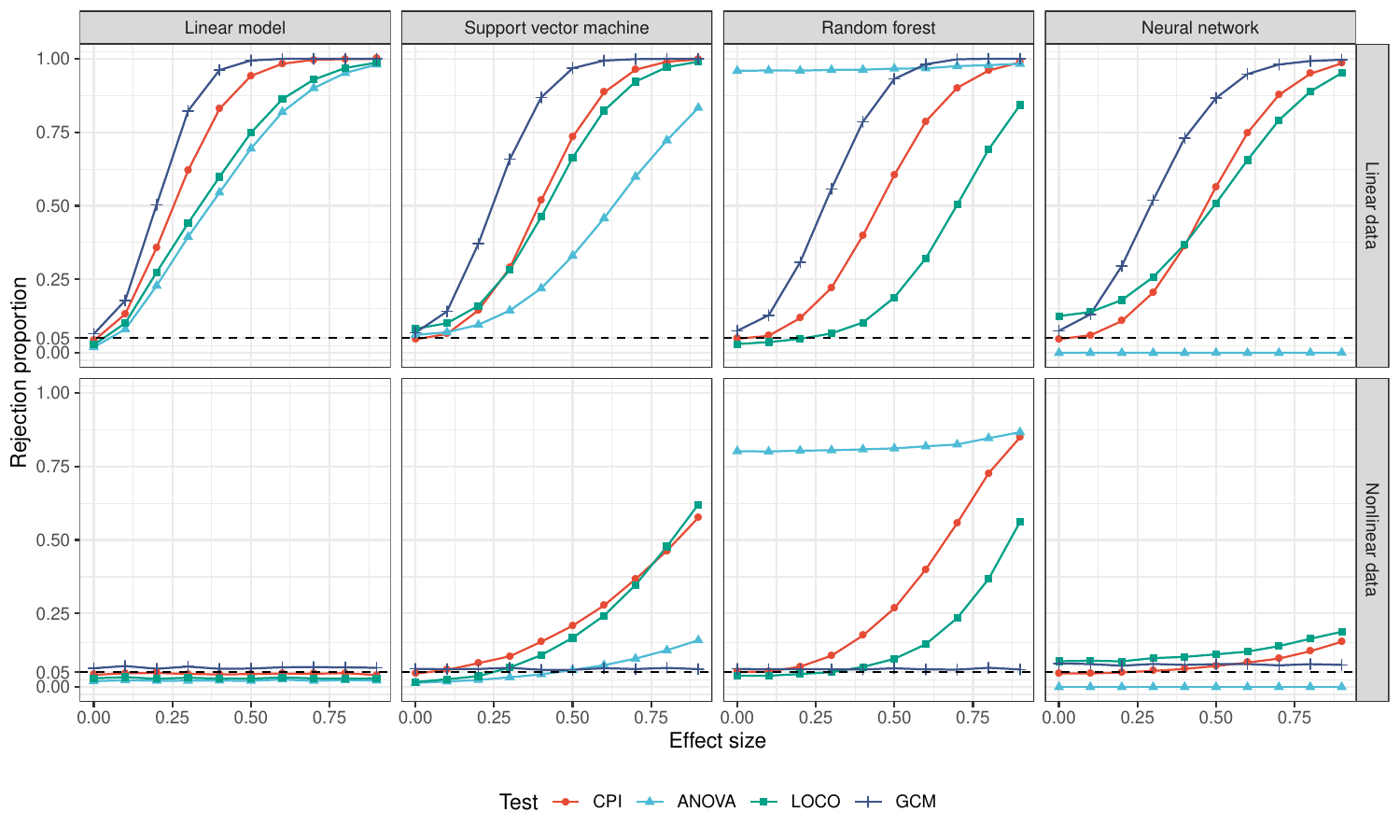}
\end{centering}
\caption{Comparative performance of VI measures across different simulations and algorithms, computed with a training and test sample of $n = 100$, $p = 10$ and uncorrelated predictors. Plots depict the proportion of rejected hypotheses at $\alpha = 0.05$ as a function of effect size. Results at effect size $0$ correspond to Type I error, at effect sizes $>0$ to statistical power. The dashed line indicates the nominal level of $\alpha = 0.05$.}
\end{figure}

\begin{figure}[htbp]
\begin{centering}
\includegraphics[width=.99\textwidth]{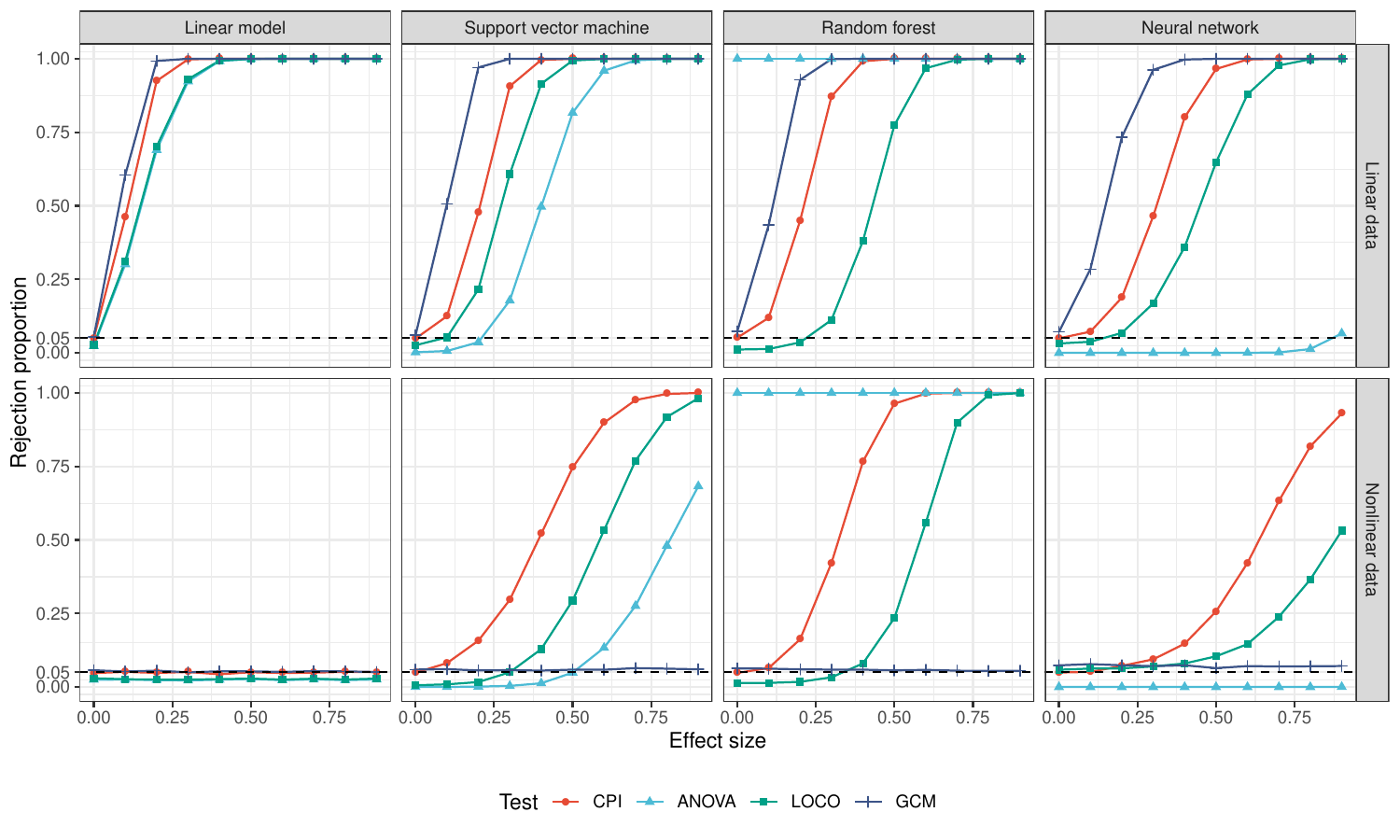}
\end{centering}
\caption{Comparative performance of VI measures across different simulations and algorithms, computed with a training and test sample of $n = 500$, $p = 10$ and uncorrelated predictors. Plots depict the proportion of rejected hypotheses at $\alpha = 0.05$ as a function of effect size. Results at effect size $0$ correspond to Type I error, at effect sizes $>0$ to statistical power. The dashed line indicates the nominal level of $\alpha = 0.05$.}
\end{figure}

\begin{figure}[htbp]
\begin{centering}
\includegraphics[width=.99\textwidth]{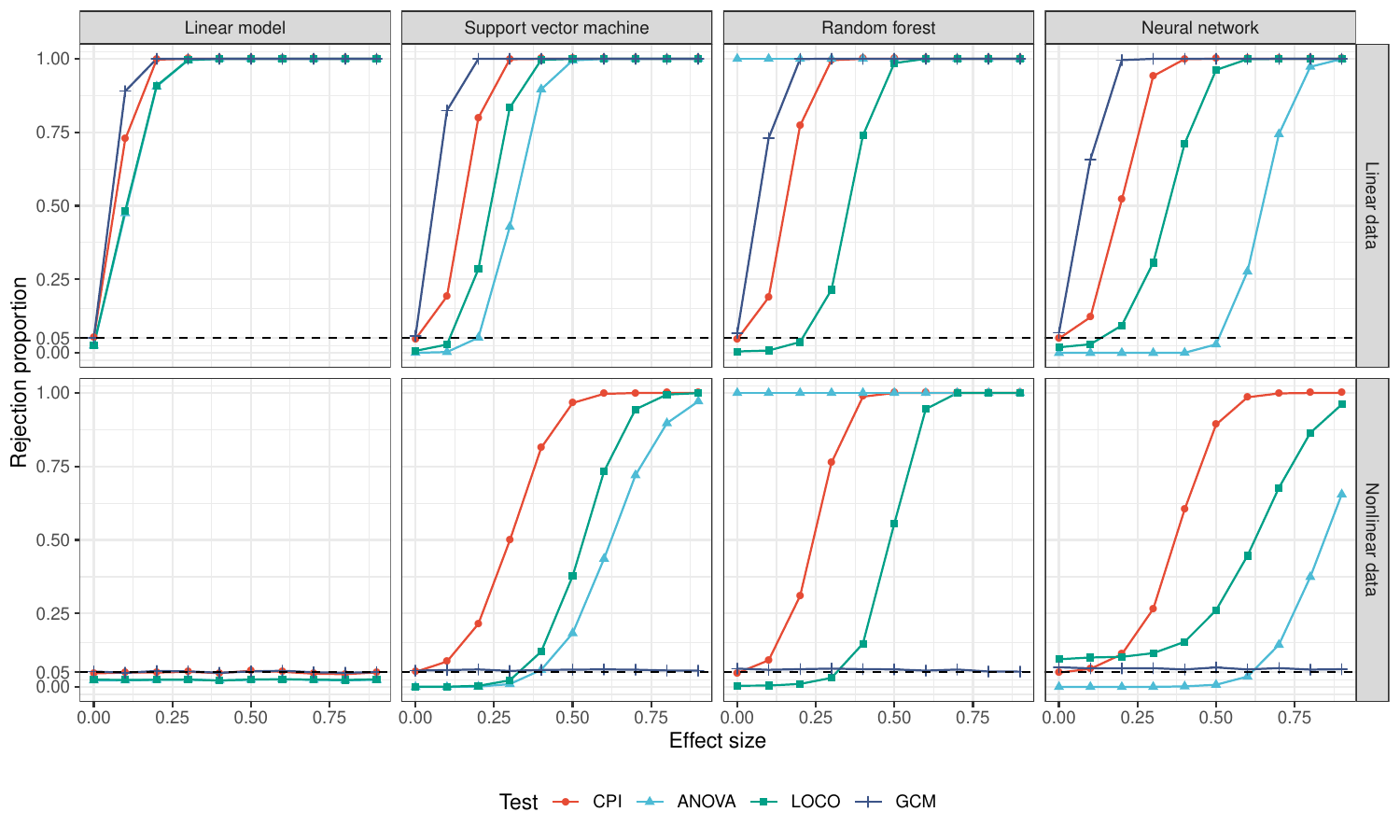}
\end{centering}
\caption{Comparative performance of VI measures across different simulations and algorithms, computed with a training and test sample of $n = 1000$, $p = 10$ and uncorrelated predictors. Plots depict the proportion of rejected hypotheses at $\alpha = 0.05$ as a function of effect size. Results at effect size $0$ correspond to Type I error, at effect sizes $>0$ to statistical power. The dashed line indicates the nominal level of $\alpha = 0.05$.}
\end{figure}

\begin{figure}[htbp]
\begin{centering}
\includegraphics[width=.99\textwidth]{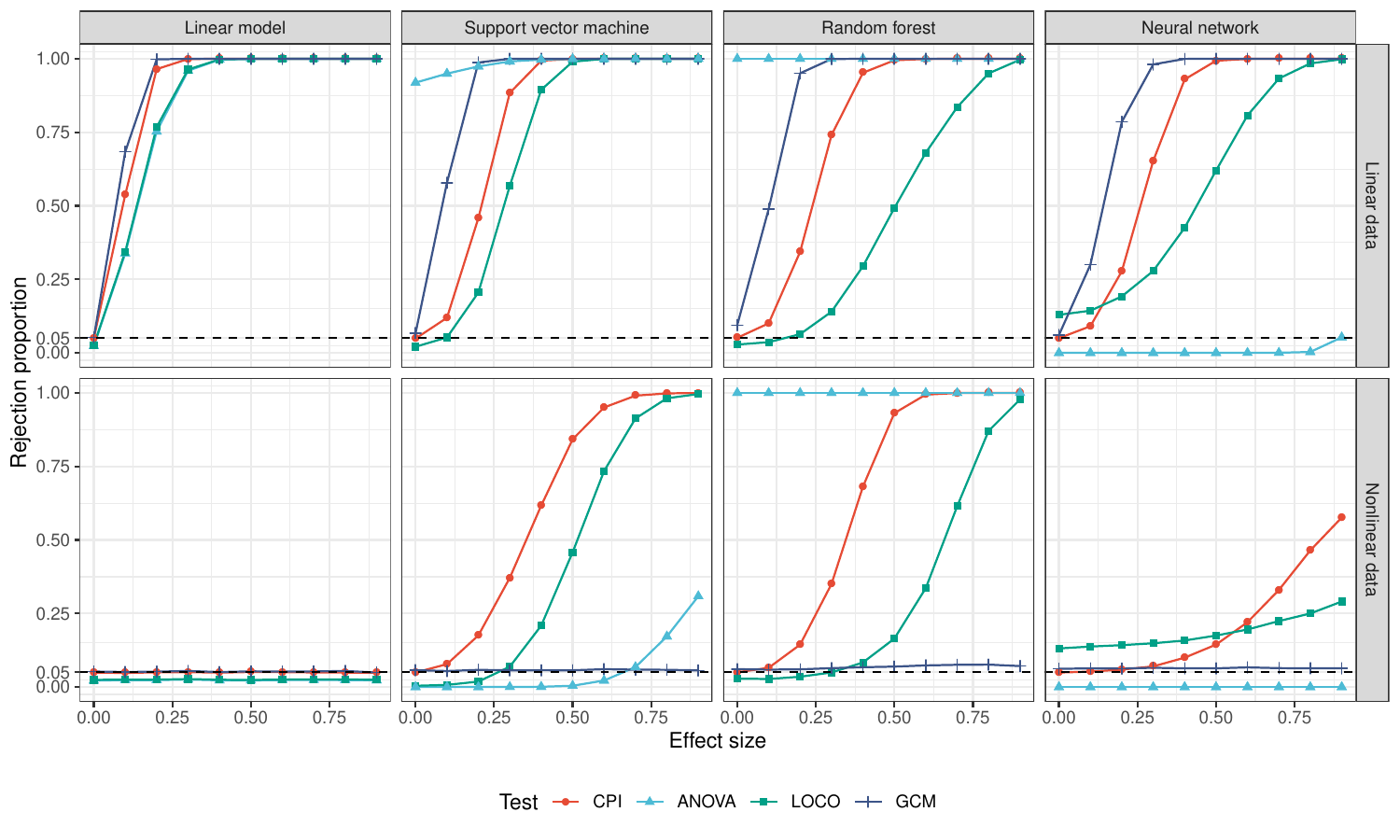}
\end{centering}
\caption{Comparative performance of VI measures across different simulations and algorithms, computed with a training and test sample of $n = 1000$, $p = 20$ and correlated predictors. Plots depict the proportion of rejected hypotheses at $\alpha = 0.05$ as a function of effect size. Results at effect size $0$ correspond to Type I error, at effect sizes $>0$ to statistical power. The dashed line indicates the nominal level of $\alpha = 0.05$.}
\end{figure}

\begin{figure}[htbp]
\begin{centering}
\includegraphics[width=.99\textwidth]{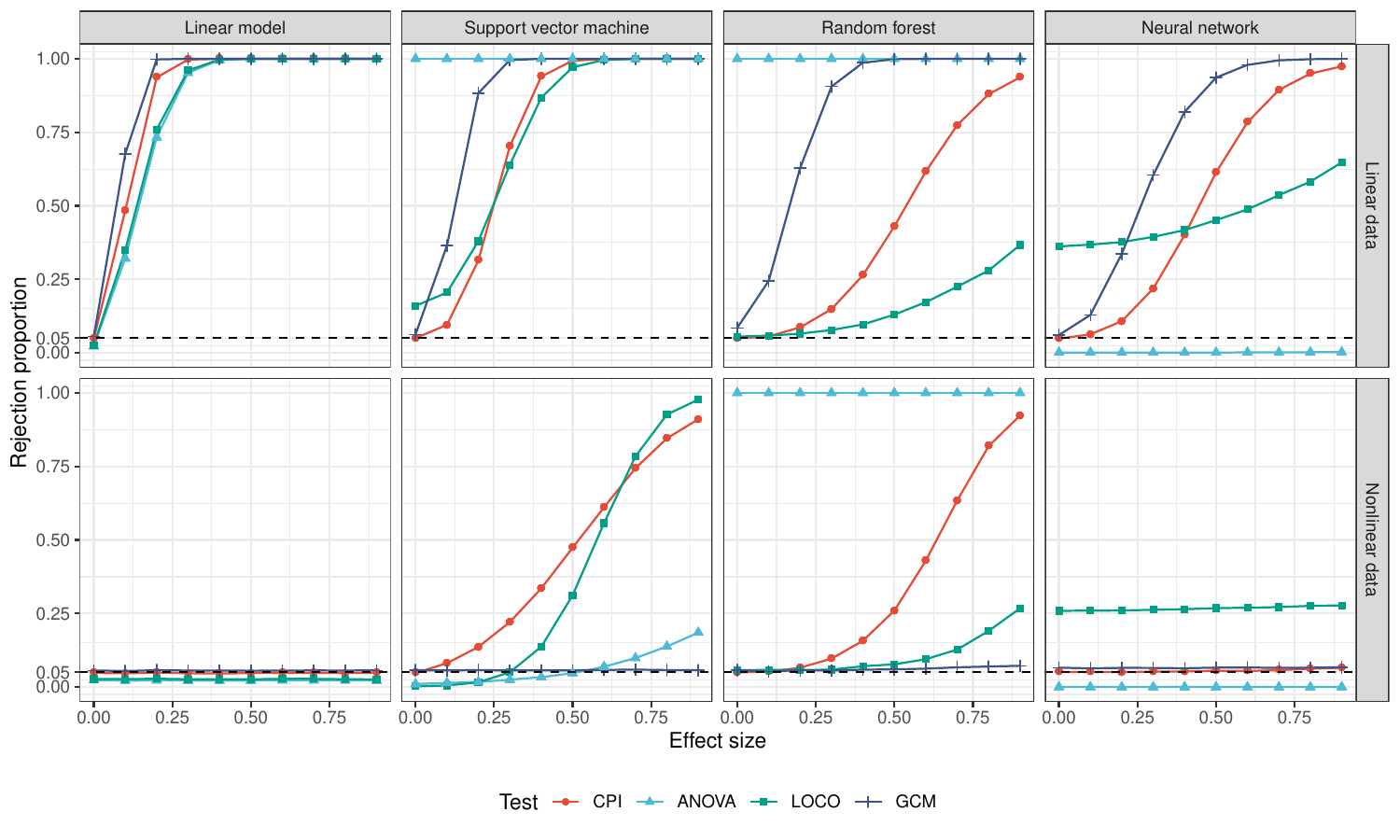}
\end{centering}
\caption{Comparative performance of VI measures across different simulations and algorithms, computed with a training and test sample of $n = 1000$, $p = 50$ and correlated predictors. Plots depict the proportion of rejected hypotheses at $\alpha = 0.05$ as a function of effect size. Results at effect size $0$ correspond to Type I error, at effect sizes $>0$ to statistical power. The dashed line indicates the nominal level of $\alpha = 0.05$.}
\end{figure}

\begin{figure}[htbp]
\begin{centering}
\includegraphics[width=.99\textwidth]{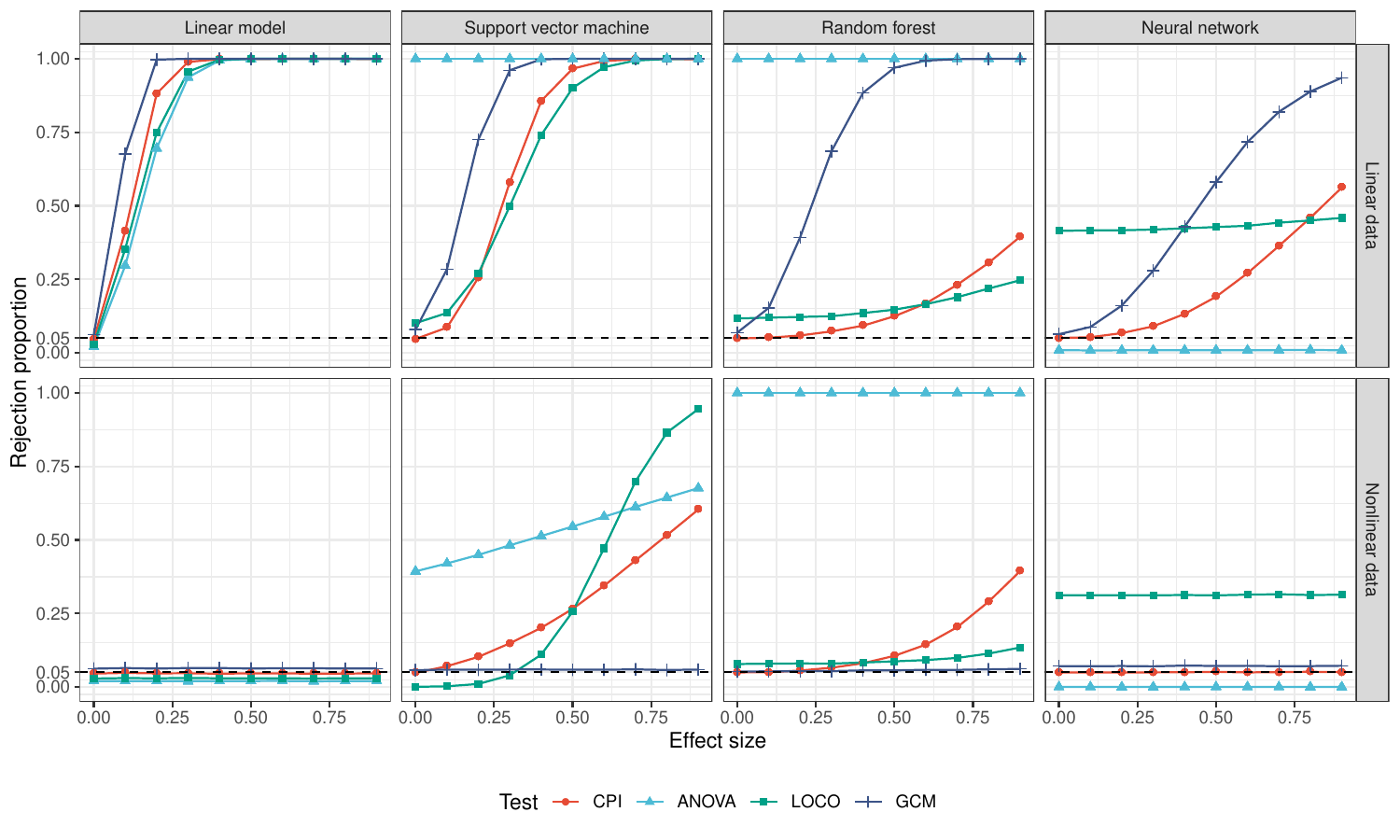}
\end{centering}
\caption{Comparative performance of VI measures across different simulations and algorithms, computed with a training and test sample of $n = 1000$, $p = 100$ and correlated predictors. Plots depict the proportion of rejected hypotheses at $\alpha = 0.05$ as a function of effect size. Results at effect size $0$ correspond to Type I error, at effect sizes $>0$ to statistical power. The dashed line indicates the nominal level of $\alpha = 0.05$.}
\end{figure}

\begin{figure}[!h]
\begin{center}
\includegraphics[width=\textwidth]{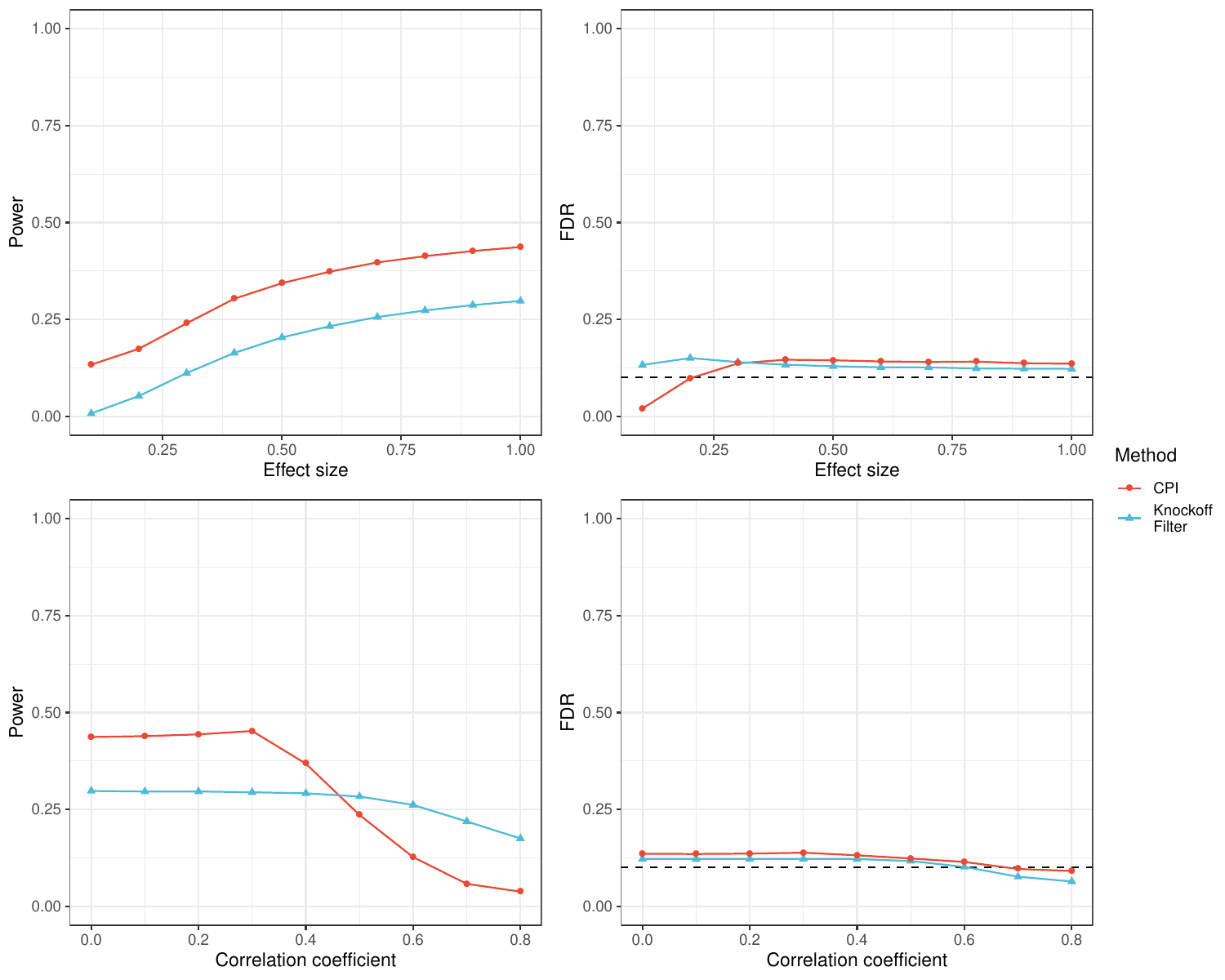}
\end{center}
\caption{Power and FDR as a function of effect size and autocorrelation for CPI and knockoff filter. Target FDR is 10\%. Results are from a lasso regression with $n = 300$ and $p = 2,000$. Each point represents 10,000 replications.}
\label{fig:kovscpi}
\end{figure}